\RequirePackage{fix-cm}
\documentclass[smallextended]{svjour3}       
\smartqed  
\usepackage{graphicx}
\usepackage{amsmath}
\usepackage{amssymb}
\usepackage{nicefrac}
\usepackage{booktabs}
\usepackage{array}

\newcommand{\rv}{\mathbf{r}}

\newcommand{\ISI}{\mathrm{ISI}}
\newcommand{\revISI}{\mathrm{revISI}}
\newcommand{\LB}{\mathrm{LB}}
\newcommand{\SPL}{\mathrm{SPL}}

\begin{document}

\title{Local and global interpolations along the adiabatic connection of DFT: A study at different correlation regimes 
\thanks{This paper is dedicated to J\'anos \'Angy\'an: his search for new understanding and passion for research continues to be a great source of inspiration.}
}

\titlerunning{Local and Global Interpolations in DFT}        

\author{Derk P. Kooi         \and
        Paola Gori-Giorgi 
}


\institute{D. P. Kooi and P. Gori-Giorgi \at
              Department of Theoretical Chemistry and Amsterdam Center for Multiscale Modeling, Faculty of Science, Vrije Universiteit, De Boelelaan 1083, 1081HV Amsterdam, The Netherlands \\
              \email{d.p.kooi@vu.nl}\\
    			\email{p.gorigiorgi@vu.nl}
}

\date{Received: date / Accepted: date}

\maketitle

\begin{abstract}
Interpolating the exchange-correlation energy along the density-fixed adiabatic connection of density functional theory is a promising way to build approximations that are not biased towards the weakly correlated regime. These interpolations can be done at the global (integrated over all spaces) or at the local level, using energy densities. Many features of the relevant energy densities as well as several different ways to construct these interpolations, including comparisons between global and local variants, are investigated here for the analytically solvable Hooke's atom series, which allows for an exploration of different correlation regimes. We also analyze different ways to define the correlation kinetic energy density, focusing on the peak in the kinetic correlation potential.
\end{abstract}

\section{Introduction}
\label{intro}
The density-fixed adiabatic connection \cite{LanPer-SSC-75} of Kohn-Sham (KS) density functional theory (DFT) is a powerful theoretical tool for the construction of approximate exchange-correlation (XC) functionals: for example, hybrid \cite{Bec-JCP-93a} and double-hybrid functionals \cite{Gri-JCP-06} can be constructed from simple models of the adiabatic connection integrand \cite{ShaTouSav-JCP-11,BreAda-JCP-11,TouShaBreAda-JCP-11}. These approximations, however, use exact ingredients only for the limit of small coupling strength, and are thus biased towards the weakly-correlated regime.

A class of approximations that removes this bias is based on the idea of Seidl and coworkers \cite{SeiPerLev-PRA-99,SeiPerKur-PRA-00,SeiPerKur-PRL-00} to interpolate the adiabatic connection integrand between its weak and strong interaction limits. This way, information from both extreme correlation regimes is taken into account on a similar footing. These interpolations can be done on the global \cite{SeiPerLev-PRA-99,SeiPerKur-PRA-00,SeiPerKur-PRL-00,FabGorSeiDel-JCTC-16,GiaGorDelFab-JCP-18,VucGorDelFab-JPCL-18} (i.e., integrated over all space) ingredients, or in each point of space, using energy densities \cite{MirSeiGor-JCTC-12,VucIroSavTeaGor-JCTC-16,VucIroWagTeaGor-PCCP-17}. As well known, energy densities are not uniquely defined and one should be sure, when doing an interpolation between weak and strong coupling in each point of space, that all the input local quantities are defined in the same way \cite{MirSeiGor-JCTC-12,VucIroSavTeaGor-JCTC-16,VucIroWagTeaGor-PCCP-17,VucLevGor-JCP-17}, which makes the use of semilocal approximations very difficult, a problem shared with local hybrids \cite{JarScuErn-JCP-03,ArbKau-CPL-07,ArbBahKau-JPCA-09,ArbKau-JCP-14}. Non-local functionals  for the strong-interaction limit \cite{WagGor-PRA-14,BahZhoErn-JCP-16} or the physical regime \cite{VucGor-JPCL-17} are needed in this context, as full compatibility with the exact exchange energy density is required.

Interpolations constructed from the global ingredients are in general computationally cheaper than their local counterpart, not only because they can use semilocal approximations for the strong-interaction functionals, but also because they do not need energy densities from exact exchange and from second-order perturbation theory, but only their global values. These global interpolations are in principle not size consistent, but it has been recently shown that their size-consistency error can be fully corrected at no additional computational cost \cite{VucGorDelFab-JPCL-18}, allowing for the calculation of meaningful interaction energies  \cite{VucGorDelFab-JPCL-18}.
On the other hand, in all the tests performed so far on small chemical systems \cite{VucIroSavTeaGor-JCTC-16,VucIroWagTeaGor-PCCP-17}, the local interpolations have always been found to be more accurate than the corresponding global ones for systems with more than two electrons. In the Helium isoelectronic series, the global and local interpolation perform similarly \cite{VucIroSavTeaGor-JCTC-16}.

The purpose of the present work is to further compare and analyze local and global interpolations when the physical system is in different correlation regimes. In order to disentangle the errors coming from the interpolation itself from those on the input ingredients, we use a model system, two Coulombically interacting electrons in the harmonic potential (``Hooke's atoms'') \cite{Tau-PRA-93,CioPer-JCP-00,MatCioVyb-PCCP-10}, which allows us to explore the whole range from weak to strong correlation always using exact input ingredients. We also analyze the kinetic correlation energy density, and particularly how its peak in the origin, which in systems with Coulomb confinement plays an important role for strong correlation \cite{BuiBaeSni-PRA-89,HelTokRub-JCP-09,YinBroLopVarGorLor-PRB-16}, varies as the system becomes more and more correlated.

\section{Theoretical Background}
\label{sec:theory}
\subsection{Density fixed adiabatic connection}\label{subsec:adiab}
By defining the $\lambda$-dependent density functional $F_{\lambda}[\rho]$ in the Levy constrained-search formalism \cite{Lev-PNAS-79}, 
\begin{equation} \label{eq:Flambda}
	F_{\lambda}[\rho]\equiv \min_{\Psi\to\rho}\langle\Psi|\hat{T}+\lambda \hat{W}|\Psi\rangle,
\end{equation}
with $\hat{T}$ the electronic kinetic energy operator, $\hat{W}$ the Coulomb electron-electron interaction operator, and ``$\Psi\to\rho$'' indicating all fermionic wavefunctions yielding the one-electron density $\rho(\rv)$, one obtains an exact formula \cite{LanPer-SSC-75} for the XC energy functional of KS DFT,
\begin{equation}\label{eq:Excadiab}
	E_{xc}[\rho]=\int_0^1 W_{\lambda}[\rho]\,d\lambda.
\end{equation}
In Eq.~(\ref{eq:Excadiab}) $W_{\lambda}[\rho]$ is the global adiabatic connection integrand,
\begin{equation}\label{eq:Wlambda}
	W_{\lambda}[\rho]\equiv \langle\Psi_{\lambda}[\rho]|\hat{W}|\Psi_{\lambda}[\rho]\rangle-U[\rho],
\end{equation}
where $\Psi_{\lambda}[\rho]$ is the minimizing wavefunction in Eq.~(\ref{eq:Flambda}) and $U[\rho]$ is the Hartree repulsion energy. The real parameter $\lambda$ is a knob that controls the interaction strength, defining an infinite set of systems all with the same one-electron density $\rho(\rv)=\rho_{\lambda=1}(\rv)$, but with different correlation. The global adiabatic connection integrand has the known expansions at small and large $\lambda$,
\begin{eqnarray}
	W_{\lambda\to 0}[\rho] & = & W_0[\rho]+\lambda\,W_0'[\rho]+...,\label{eq_weak}\\
	W_{\lambda\to \infty}[\rho] & = & W_\infty[\rho]+\frac{W'_\infty[\rho]}{\sqrt{\lambda}}+... \label{eq_strong},
\end{eqnarray}
where $W_0[\rho]=E_x[\rho]$ is the exact exchange energy (the same expression as the Hartree-Fock exchange, but with KS orbitals), $W_0'[\rho]=2 E_c^{\rm GL2}[\rho]$ is twice the G\"orling-Levy \cite{GorLev-PRA-94} second-order correlation energy (GL2), $W_\infty[\rho]$ is the indirect part of the minimum possible expectation value of the electron-electron repulsion in a given density \cite{SeiGorSav-PRA-07}, and $W'_\infty[\rho]$ is the potential energy of coupled zero-point oscillations around the manifold that determines $W_\infty[\rho]$ \cite{GorVigSei-JCTC-09}.

\subsection{Energy densities}
\label{subsec:endens}
Equation~(\ref{eq:Excadiab}) can also be written in terms of real-space energy densities $w_\lambda(\rv;[\rho])$,
\begin{equation}
\label{eq:energydensity}
 E_{xc}[\rho] = \int d \mathbf{r} \, \rho(\mathbf{r})\int_0^1 \text{d} \lambda\,  w_\lambda(\mathbf{r}; [\rho]),
\end{equation}
which are, of course, not uniquely defined. For the purpose of building $\lambda$-interpolation models on energy densities, the choice of the gauge of the electrostatic potential of the exchange-correlation hole $h^\lambda_{xc}(\mathbf{r}_1, \mathbf{r}_2)$ seems so far to be the most suitable \cite{VucLevGor-JCP-17},
\begin{equation}
 w_\lambda(\mathbf{r}) = \frac{1}{2} \int \frac{h^\lambda_{xc}(\mathbf{r}, \mathbf{r}_2)}{|\mathbf{r}-\mathbf{r}_2|} d \mathbf{r}_2,
 \label{eq:energydef}
\end{equation}
 where $h^\lambda_{xc}(\mathbf{r}_1, \mathbf{r}_2)$ is defined in terms of the pair-density $P_2^\lambda(\mathbf{r}_1, \mathbf{r}_2)$ and the density $\rho$ (see also \cite{GorAngSav-CJC-09}),
\begin{equation}
 h^\lambda_{xc}(\mathbf{r}_1, \mathbf{r}_2) = \frac{P_2^\lambda(\mathbf{r}_1, \mathbf{r}_2)}{\rho(\mathbf{r}_1)} - \rho(\mathbf{r}_2),
\end{equation}
 with $P_2^\lambda$ obtained from $\Psi_\lambda [\rho]$,
\begin{equation}
 P_2^\lambda(\mathbf{r}, \mathbf{r}') = N(N-1) \sum_{\sigma, \sigma', \sigma_3 \dots \sigma_N} \int | \Psi_\lambda(\mathbf{r}\sigma, \mathbf{r}'\sigma', \mathbf{r}_3\sigma_3 \dots r_N \sigma_N)|^2 d \mathbf{r}_3 \dots d \mathbf{r}_N.
 \label{pddef}
\end{equation}
\paragraph{Energy density at $\lambda = 0$.}
At $\lambda = 0$ we have the Kohn-Sham or exchange hole, which yields, in the case of a closed-shell singlet considered in this work (with real orbitals)
\begin{equation}  \label{eq:w0cs}
w_0(\mathbf{r}) = - \frac{1}{2 \rho(\mathbf{r})} \sum_{i, j}^{N/2}  \phi_i(\mathbf{r}) \phi_j(\mathbf{r}) \int \mathrm{d} \mathbf{r}' \frac{\phi_j(\mathbf{r}') \phi_i(\mathbf{r}')}{|\mathbf{r}-\mathbf{r}'|},
\end{equation}
where $\phi_i(\mathbf{r})$ are the occupied KS spatial orbitals.

\paragraph{Slope of the energy density at $\lambda = 0$.}
 The slope $w'_0(\mathbf{r})$ of the energy density at $\lambda = 0$  in the gauge of Eq.~(\ref{eq:energydef}) is given, again for a closed shell singlet with real orbitals, by \cite{VucIroSavTeaGor-JCTC-16}
\begin{align} \label{eq:w0primecs}
 w'_0(\mathbf{r}) = - \frac{1}{\rho(\mathbf{r})} \sum_{abij} \frac{4 \langle ij | ab \rangle -2 \langle ij | ba \rangle}{\epsilon_a + \epsilon_b - \epsilon_i - \epsilon_j}  \phi_i(\mathbf{r}) \phi_a(\mathbf{r}) \int d \mathbf{r}' \frac{\phi_j(\mathbf{r}') \phi_b(\mathbf{r}') }{|\mathbf{r} - \mathbf{r}'|},
\end{align}
where $\phi_a$ and $\phi_b$ are unoccupied and $\phi_i$ and $\phi_j$ are occupied Kohn-Sham orbitals, $\langle ij | ab \rangle$ denotes the Coulomb integral over the spatial orbitals, and the $\epsilon_i$ are the Kohn-Sham orbital energies. For systems with $N>2$, there should be also a term with single excitations \cite{GorLev-PRA-94}, which we do not consider here as we focus on $N=2$.
 
\paragraph{Energy density at $\lambda = \infty$.} \label{subsec:winf}
In the $\lambda\to\infty$ limit we obtain a system of strictly correlated electrons (SCE), for which it has been shown \cite{MirSeiGor-JCTC-12} that 
\begin{equation} \label{eq:winf}
w_\infty(\mathbf{r}) = \frac{1}{2} \sum_{i=2}^N \frac{1}{|\mathbf{r}-\mathbf{f}_i(\mathbf{r})|} - \frac{1}{2} v_H(\mathbf{r}),
\end{equation}
where $ v_H(\mathbf{r})$ is the Hartree potential and $\mathbf{f}_i(\mathbf{r})$ are {\em co-motion functions} that determine the position of the $i^{\rm th}$ electron given the position $\rv$ of a chosen reference electron (as the $\mathbf{f}_i(\mathbf{r})$ satisfy cyclic group properties it does not matter which electron is chosen as reference), and are non-local functionals of the density $\rho(\rv)$ \cite{SeiGorSav-PRA-07,MalMirCreReiGor-PRB-13}. 

There is at present no local expression in the gauge of Eq.~(\ref{eq:energydef}) for the next leading term $W_\infty'[\rho]$ in the $\lambda\to\infty$ asymptotic expansion. In fact, the functional $W_\infty'[\rho]$ can be computed from an integral on position-dependent zero-point energies \cite{GorVigSei-JCTC-09}, which, however, do not provide an energy density within the definition of  Eq.~(\ref{eq:energydef}).

\subsection{Global and local interpolations}
\label{subsec:interpolations}
The original idea of Seidl and coworkers \cite{SeiPerLev-PRA-99,SeiPerKur-PRA-00,SeiPerKur-PRL-00} was to build an approximate adiabatic connection integrand $W_\lambda^{\rm ISI}[\rho]$ by interpolating between the two limits of Eqs.~(\ref{eq_weak}) and (\ref{eq_strong}). These interaction-strength interpolation (ISI) functionals typically use as input the four ingredients (or a subset thereof) appearing in Eqs.~(\ref{eq_weak}) and (\ref{eq_strong}): $\{W_0[\rho],W_0'[\rho], 
W_\infty[\rho], W_\infty'[\rho]\}$, denoted ${\bf W}$ in short. The XC energy functional $E_{xc}^{\rm ISI}[\rho]$ is then obtained from Eq.~(\ref{eq:Excadiab}), by integrating $W_\lambda^{\rm ISI}[\rho]$ over $\lambda$, which will result in a non-linear function of the input ingredients ${\bf W}$. Because of this non linear dependence, the ISI-type functionals are not size consistent when a system dissociates into unequal fragments, even when the input ingredients are size-consistent themselves. However, in this latter case, size-consistency can be easily restored with a very simple correction \cite{VucGorDelFab-JPCL-18}. The ISI-type functionals are, instead, automatically size extensive \cite{VucGorDelFab-JPCL-18}.
Several formulas for interpolating between the two limits of Eqs.~(\ref{eq_weak}) and (\ref{eq_strong}) have been proposed in the literature, and are reported in Appendix~\ref{app:formulas}. 

More recently, these same interpolation formulas have been used to build,  in each point of space, a model energy density $w_\lambda^{\rm ISI}(\rv;[\rho])$, with Eqs.~(\ref{eq:w0cs})-(\ref{eq:winf}) as input ingredients \cite{VucIroSavTeaGor-JCTC-16,VucIroWagTeaGor-PCCP-17}. This way, by integrating $w_\lambda^{\rm ISI}(\rv;[\rho])$ over $\lambda$ between 0 and 1, one obtains an exchange-correlation energy density in the gauge of the coupling-constant averaged exchange-correlation hole. Such interpolations done in each point of space are size consistent in the usual DFT sense \cite{GorSav-JPCS-08,Sav-CP-09}.

\subsection{Hooke's atom series}
The Hooke's atom series consists of two electrons bound by an harmonic external potential, with hamiltonian
\begin{equation}
	\label{eq:HamHooke}
	\hat{H}= - \frac{1}{2}\left(\nabla_1^2 + \nabla_2^2\right) + \frac{\omega^2}{2} \left(r_1^2 + r_2^2\right) + \frac{1}{r_{12}},
\end{equation}
with $r_i=|\mathbf{r}_i|$ and $r_{12}=|\rv_1-\rv_2|$.
At large $\omega$ the system has high-density and is in the weakly correlated regime, which can be fully described by using the scaled coordinates $\mathbf{s}_i\equiv \sqrt{\omega}\,\rv_i$, while as $\omega\to 0$ the system becomes more and more correlated \cite{CioPer-JCP-00}, and the relevant scaled variables are $\tilde{\mathbf{s}}_i\equiv \omega^{2/3}\,\rv_i$.

As well known, there is an infinite set of special values of $\omega$ for which the hamiltonian (\ref{eq:HamHooke}) is analytically solvable \cite{Tau-PRA-93} once rewritten in terms of center of mass and relative coordinates. These analytic solutions have the center of mass in the ground-state of an harmonic oscillator with mass $m=2$ and frequency $\sqrt{2}\,\omega$, and the relative coordinate in an $s$-wave with the radial part described by a gaussian times a polynomial \cite{Tau-PRA-93}. We denote here the various analytic solutions with the degree $n-1$ of the polynomial in $r_{12}$. At $n=1$ we have the non-interacting system, and as $n$ increases the system becomes more and more correlated, with $\omega$ smaller and smaller \cite{Tau-PRA-93}. The values of $\omega$ corresponding to the different values of $n$ considered here are reported in Table~\ref{tab:omega}.

\begin{table}
\caption{Values of $\omega$ for the various analytic solutions of the hamiltonian of Eq.~(\ref{eq:HamHooke}) considered here, corresponding to different degrees $n-1$ of the polynomial in the solution for the relative coordinate $r_{12}$ \cite{Tau-PRA-93}.}
\label{tab:omega}       
\begin{tabular}{ll}
\hline\noalign{\smallskip}
$n$ & $\omega$  \\
\noalign{\smallskip}\hline\noalign{\smallskip}
2 & $0.5$  \\
3 & $0.1$  \\
4 & $0.0365373$  \\
5 & $0.0173462$  \\
6 & $0.00957843$  \\
\noalign{\smallskip}\hline
\end{tabular}
\end{table}

\section{Computation of exact energy densities}
Given the analytic solutions \cite{Tau-PRA-93} $\Psi(r_1,r_2,r_{12})$ of the hamiltonian (\ref{eq:HamHooke}) for $n=2,\dots,6$, we have computed the corresponding densities $\rho(r)$, which are also analytic. Although leading to cumbersome expressions, these densities allowed us to obtain analytic Kohn-Sham potentials $v_s(r)=\frac{\nabla^2\sqrt{\rho(r)}}{2\sqrt{\rho(r)}}+\epsilon$, with $\epsilon=E_2-E_1$, the energy difference between the physical state with two and one electrons.

\subsection{Energy densities at $\lambda=0$}\label{sec:w0Hooke}
\begin{figure*}
\includegraphics[width=0.5\textwidth]{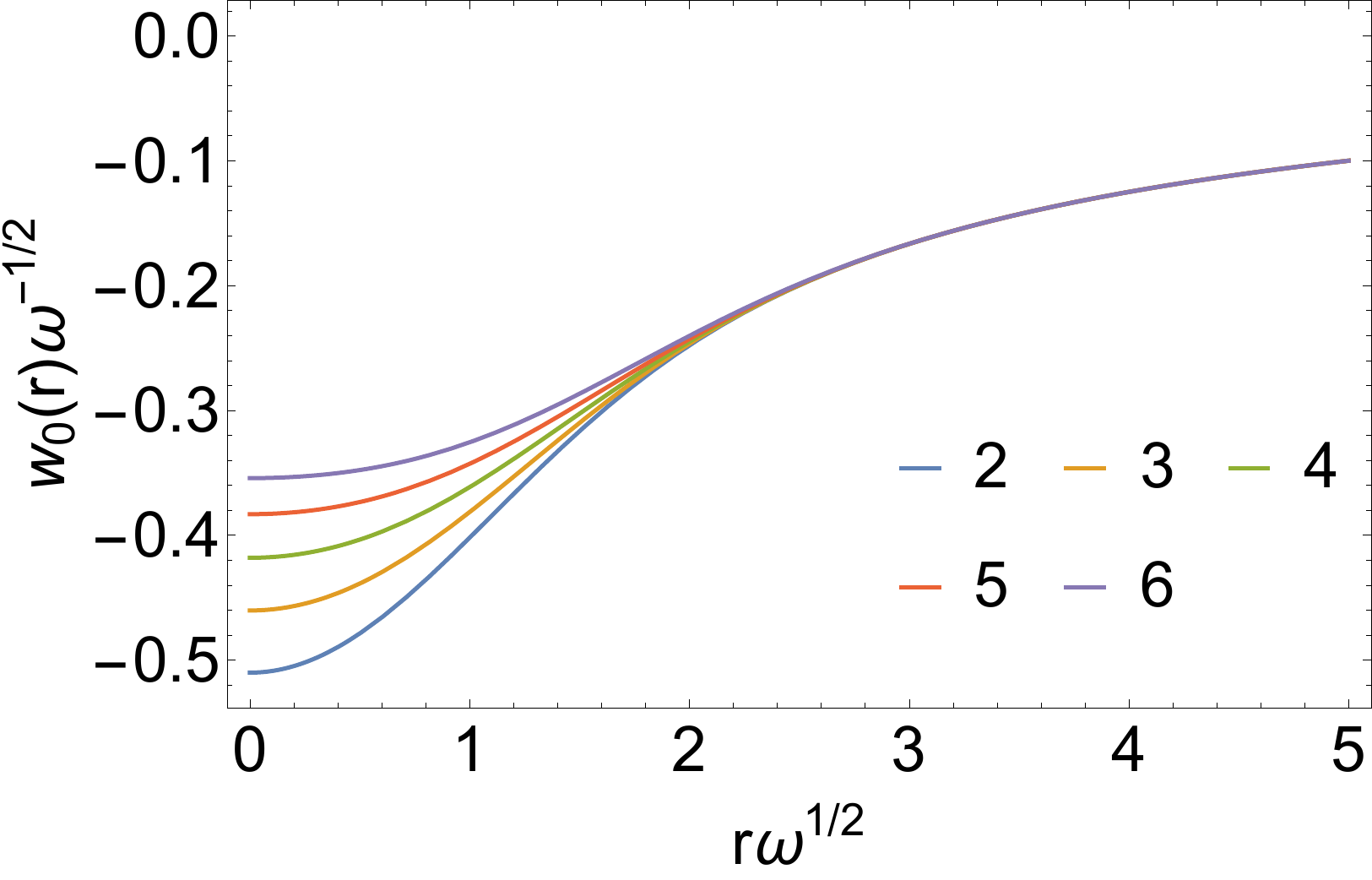}
\includegraphics[width=0.5\textwidth]{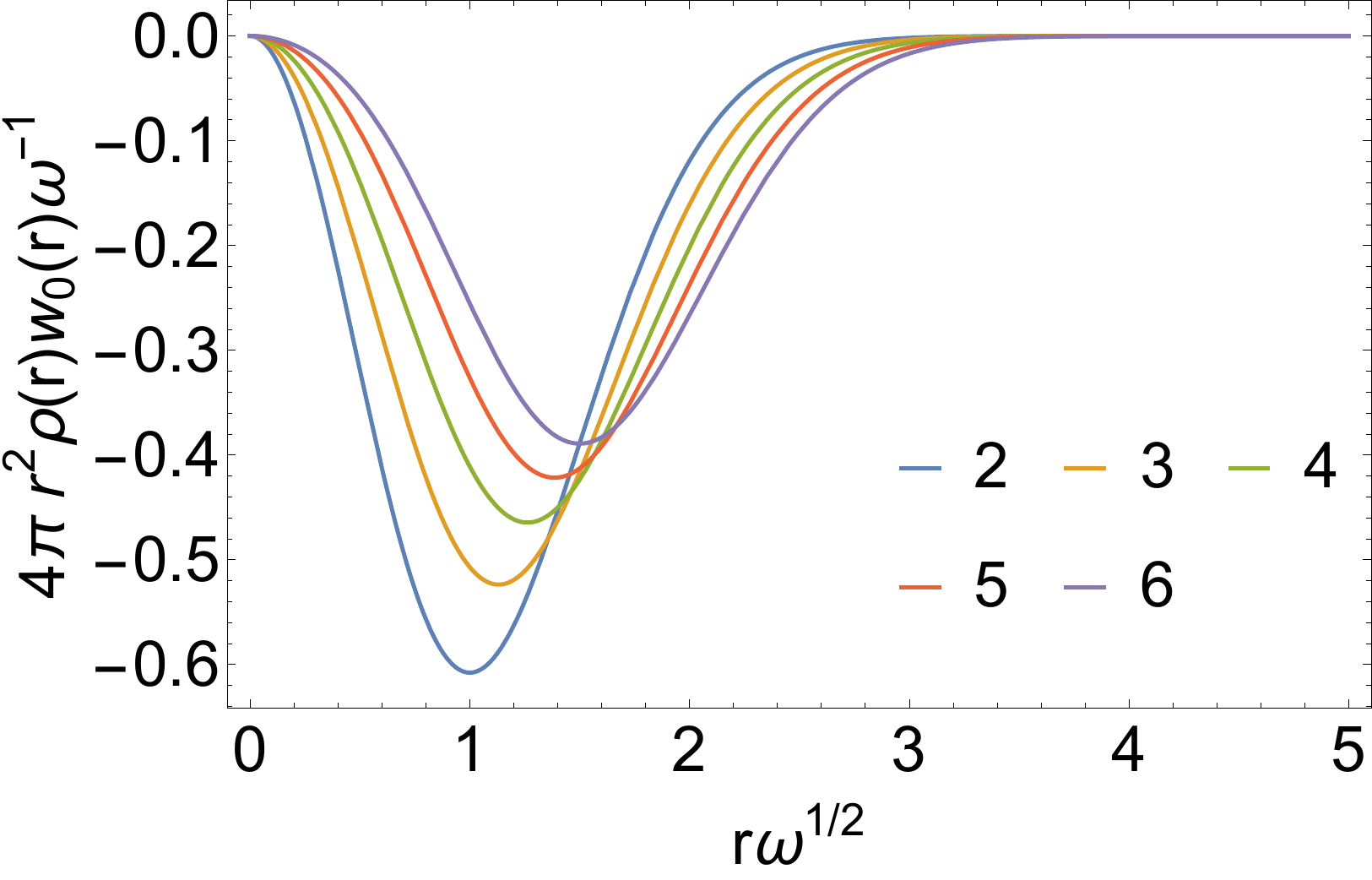}
\caption{Energy densities at $\lambda=0$ for the Hooke's atoms series with $n=2,\dots,6$, corresponding to the $\omega$ values of Table~\ref{tab:omega}. In the second panel the energy density has been multiplied by the density and by the volume element. The high-density scaling has been used.}
\label{fig:w0}       
\end{figure*}

For a singlet $N=2$ state Eq.~(\ref{eq:w0cs}) reduces to $w_0(r)=-\frac{1}{4}v_H(r)$, with $v_H(r)$ the Hartree potential, leading to the simple expression
\begin{equation}\label{eq:w0Hooke}
	w_0(r)=-\pi\int_r^\infty r'\rho(r')\,dr'-\frac{N_e(r)}{4\, r},
\end{equation}
with the cumulant $N_e(r)$ defined as
\begin{equation} \label{eq:cumulant}
	N_e(r)=4 \pi\int_0^r r'^2\rho(r')\,dr'.
\end{equation}
We have obtained these energy densities analytically from the exact densities. They are shown in Fig.~\ref{fig:w0} for the different analytic solutions considered here.

\subsection{Energy densities for the slope at $\lambda=0$} \label{sec:slopeHooke}
The analytic exact Kohn-Sham potentials were used to obtain the virtual Kohn-Sham orbitals needed for the evaluation of Eq.~(\ref{eq:w0primecs}). We used an isotropic spherical Gaussian basis with $\omega$ as the width parameter. Angular momentum values were included from $l=0$ to $l=9$, with 5 to 30 basis states for every value of $l$. All matrix elements were obtained analytically in this basis, including the Coulomb integrals. 

\begin{figure*}
  \includegraphics[width=0.5\textwidth]{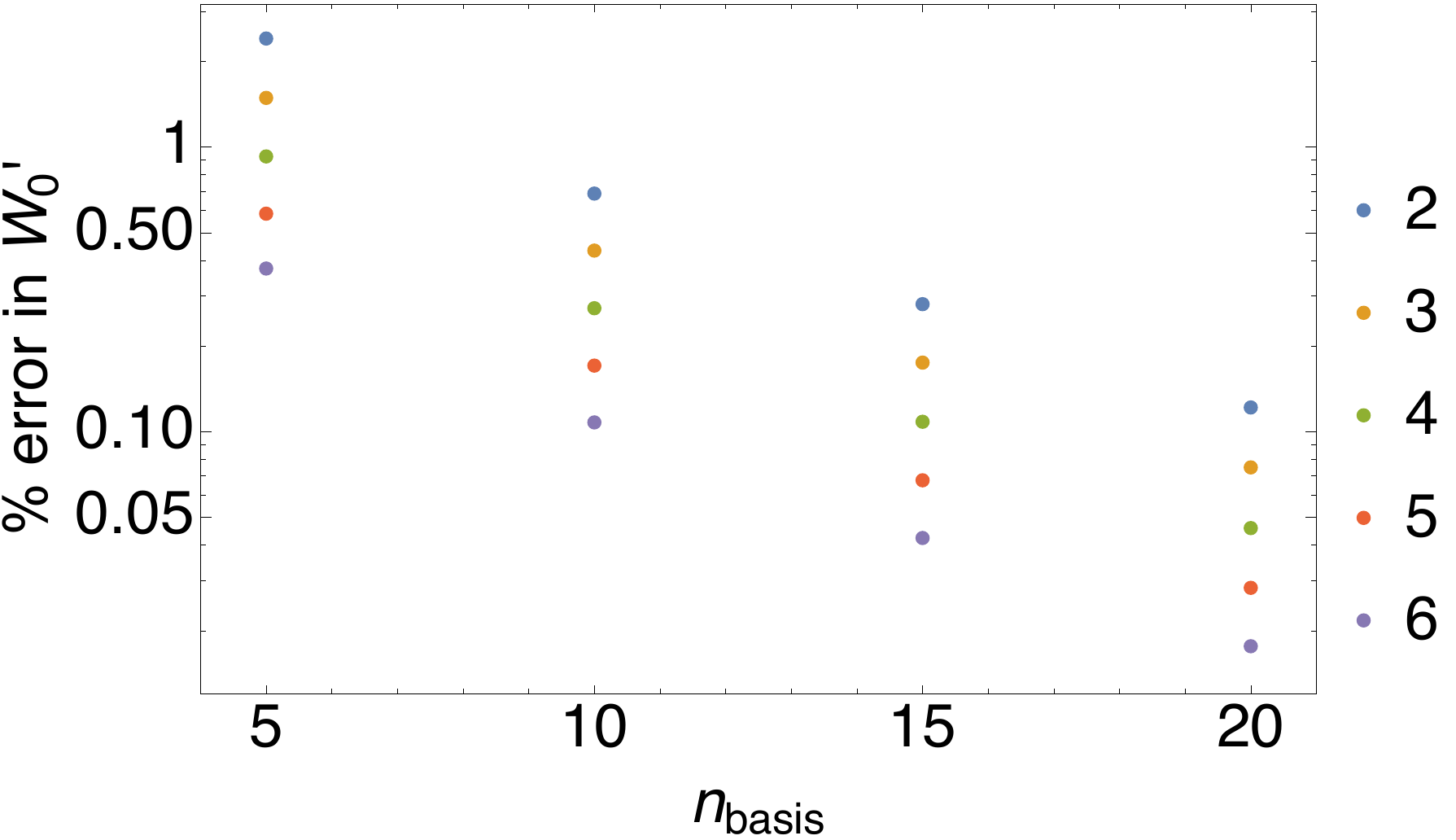}
\includegraphics[width=0.5\textwidth]{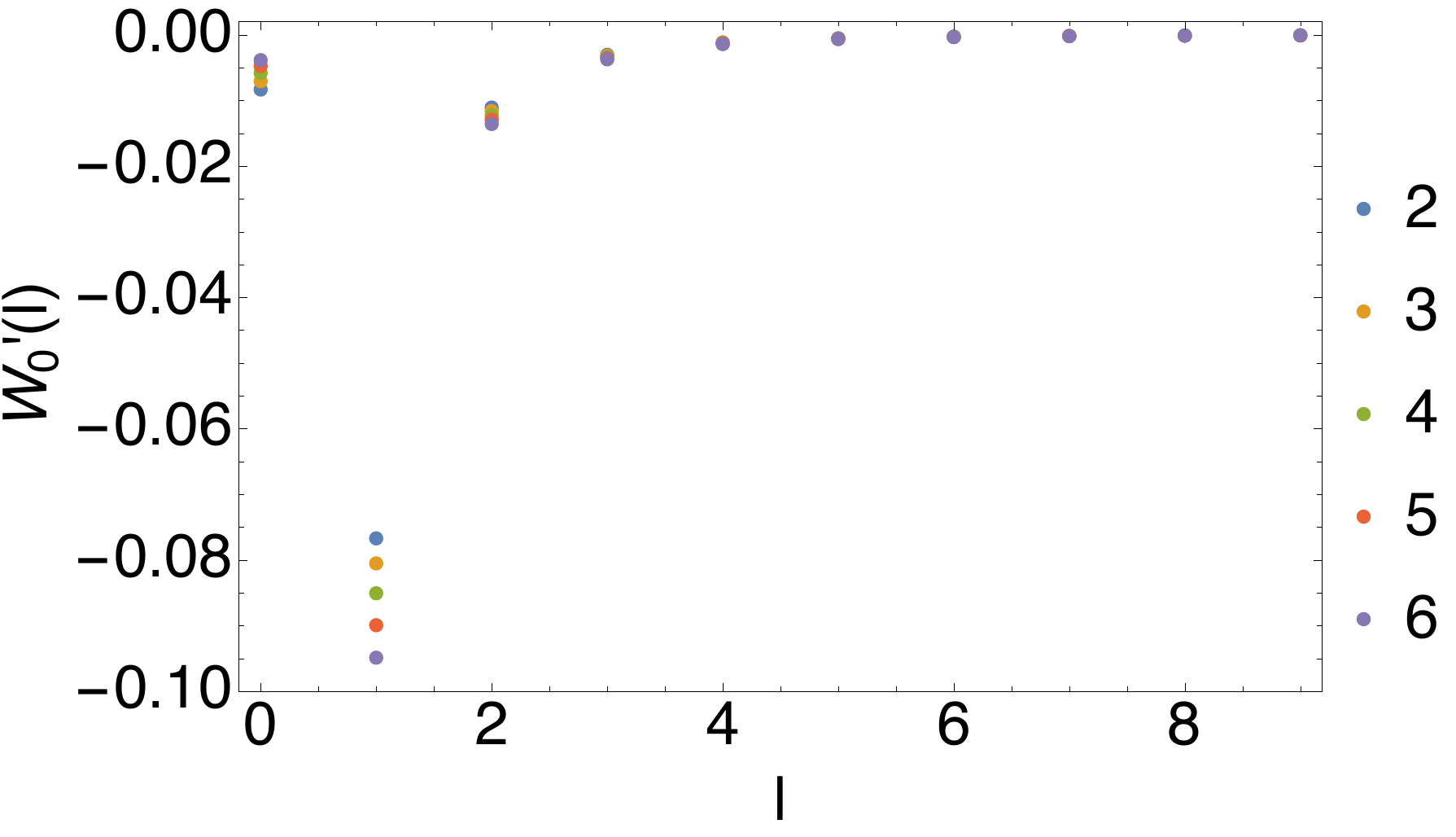}
\caption{Convergence of $W_0'=2\,E_c^{\rm GL2}$ with the size $n_{\rm basis}$ of the gaussian basis set used to expand the KS orbitals, relative to $n_{\text{basis}}=30$, (first panel) and contribution of the different angular momentum $l$ (second panel) }
\label{fig:Wd0convplot}       
\end{figure*}

We first analyze the convergence of the global slope of the coupling constant integrand, $W_0'=2\,E_c^{\rm GL2}$, with increasing basis set size $n_{\rm basis}$ in the first panel of Fig.~\ref{fig:Wd0convplot}. The number of basis states is that per angular momentum quantum number, with all $l$ up to $l=9$ included.  As $\omega$ decreases (the quantum number $n$ increases), the $l=0$ contribution becomes less important, with the $l>0$ contributions gaining more weight, as shown in the second panel of Fig.~\ref{fig:Wd0convplot}, where the result from each channel $l$ with $n_{\rm basis}=30$ is reported.

\begin{figure*}
 \includegraphics[width=0.5\textwidth]{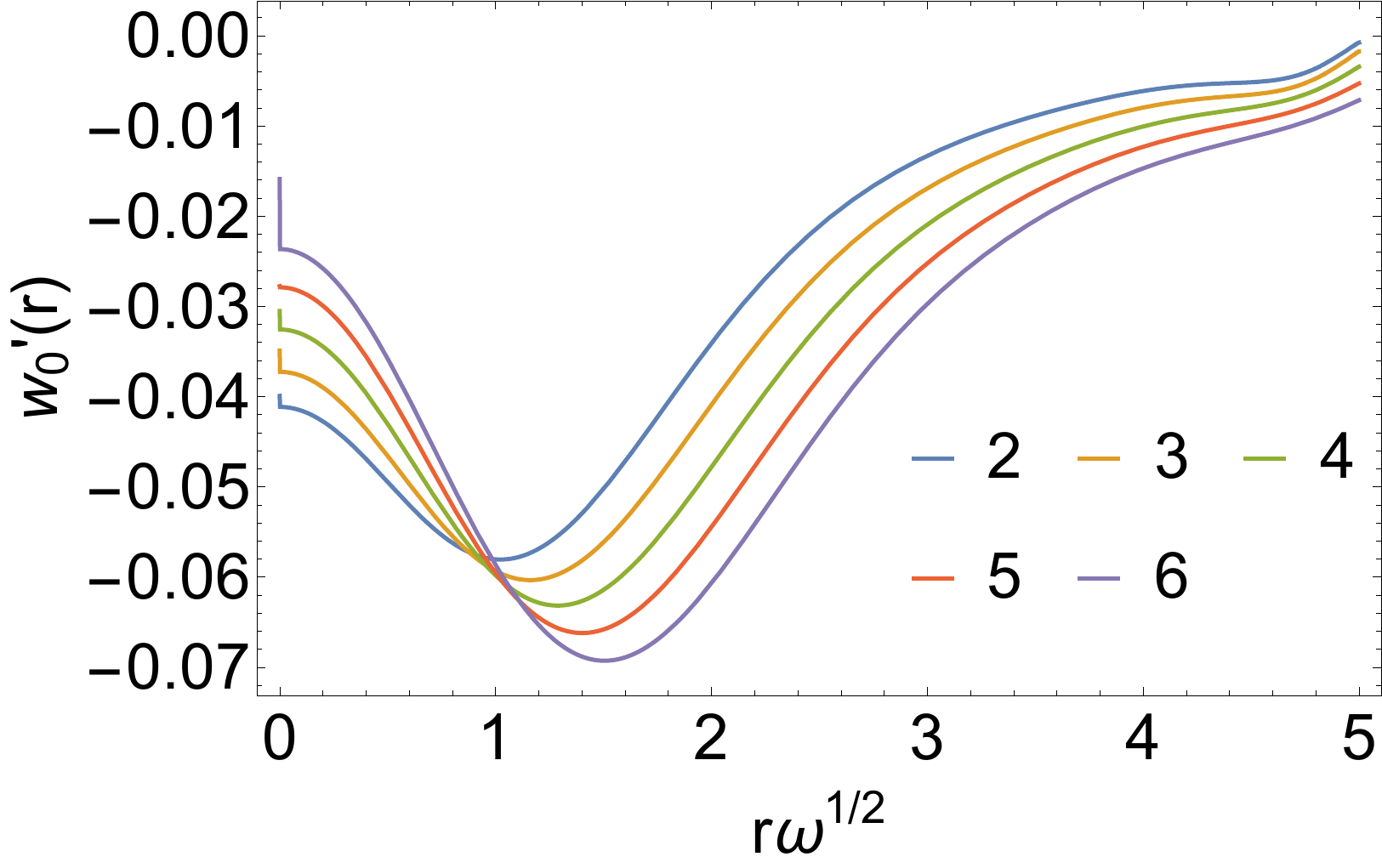}
 \includegraphics[width=0.45\textwidth]{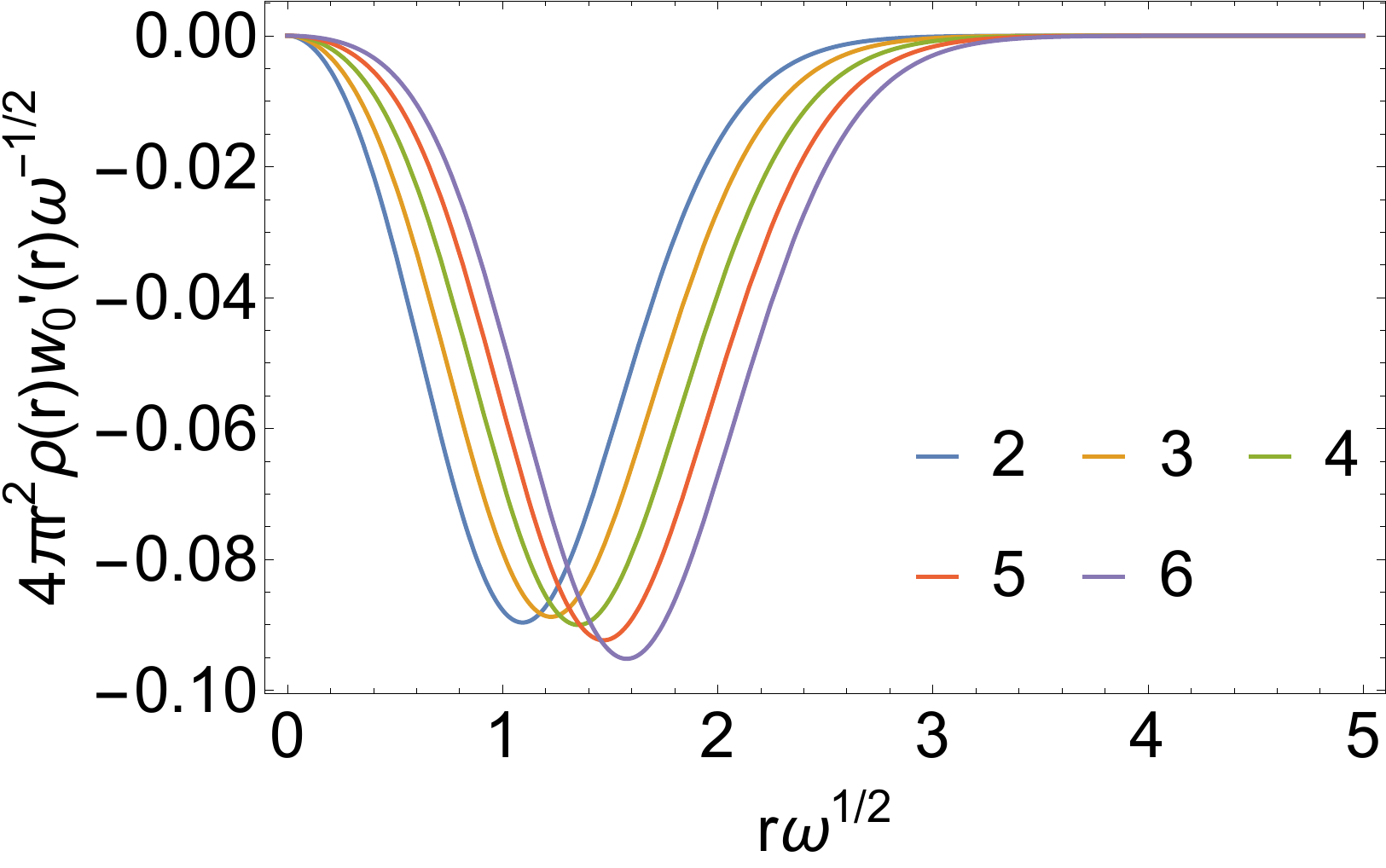}
 \caption{The local slope (first panel) and the local slope multiplied by the volume element and density (second panel).}
  \label{fig:localslope}
\end{figure*}

For the local slope $w_0'(\mathbf{r})$ only 10 basis states are used.  In the present case of a two-electron system, $w_0'(\mathbf{r})$ can also be simplified, as there is only one occupied Kohn-Sham spatial orbital. Additional utilization of the spherical symmetry then yields the following expression, by using the spherical harmonic expansion of the Coulomb potential,
\begin{equation}
\begin{split}
\label{eq:sphericalslope}
 w'_0(r) = - \frac{2}{\rho(r)} \sum_{n_an_b l} \frac{1}{\epsilon_a + \epsilon_b - 2\epsilon_{occ}} \langle (occ)(occ) | ab \rangle \\  R_{occ}^{0}(r) R_{n_a}^{l}(r) (r^{-l-1}\int_0^r dr' r'^{l +2} R_{occ}^{0}(r') R_{n_b}^{l}(r') \\+ r^l \int_r^\infty dr' r'^{-l+1} R_{n_j}^{0}(r') R_{n_b}^{l}(r')),
\end{split}
\end{equation}
where the functions $R_n^l(r)$ are the radial functions of the spatial orbitals and $occ$ is the occupied Kohn-Sham orbital. The full local slope is shown in the first panel of Fig.~\ref{fig:localslope}. Numerical issues appear at around  the scaled variable values $s \gtrsim 4.5$, but this is of no relevance to the integrated energy as it is clear upon multiplication by the volume element and the density (second panel of Fig.~\ref{fig:localslope}). 

\subsection{Energy densities at $\lambda=\infty$}
\begin{figure*}
  \includegraphics[width=0.5\textwidth]{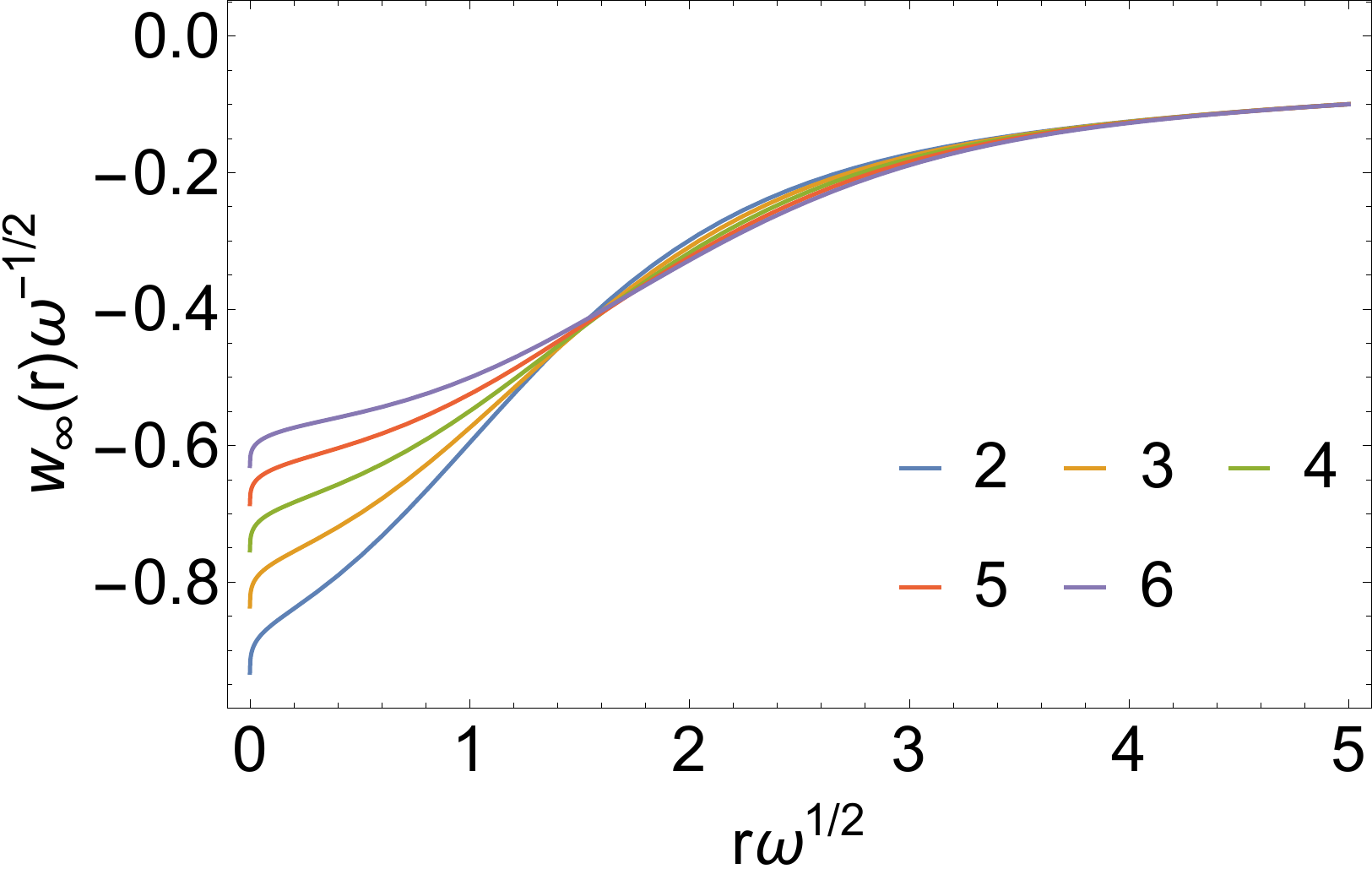}
\includegraphics[width=0.5\textwidth]{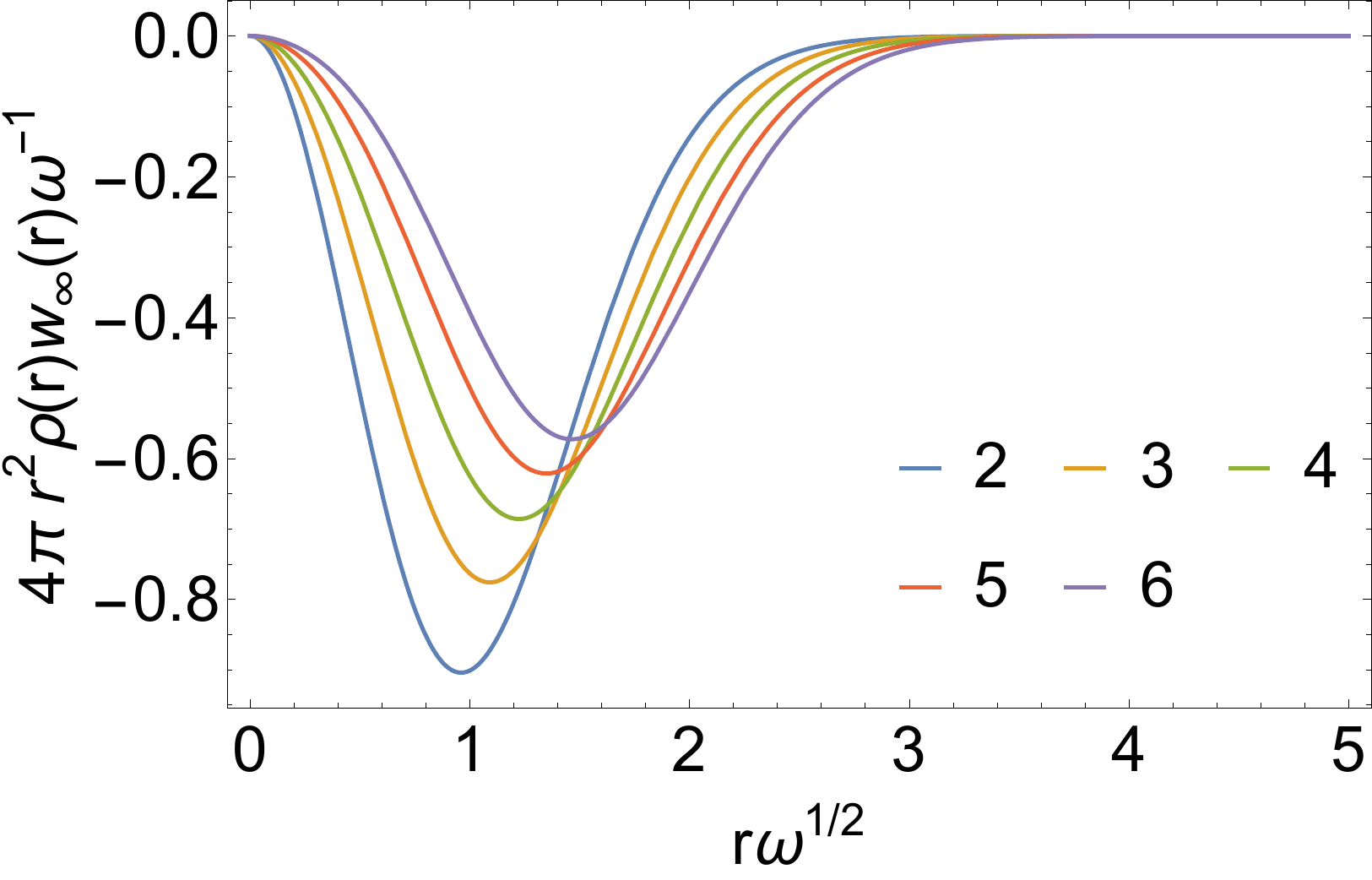}
\caption{Energy densities corresponding to $\lambda=\infty$ (first panel), and energy densities corresponding to $\lambda=\infty$ multiplied by the density and the volume element (second panel). The coordinates and energy densities are scaled according to the large $\omega$ limit.}
\label{fig:winfinity}       
\end{figure*}

The energy density $w_\infty(\mathbf{r})$ of Eq.~(\ref{eq:winf}) in the case of $N=2$ electrons in a spherical density is known to be determined by the radial co-motion function $f(r)$, which gives the full $\mathbf{f}(\mathbf{r})$ via $\mathbf{f}(\mathbf{r})=-\frac{f(r)}{r}\,\rv$  \cite{Sei-PRA-99,SeiGorSav-PRA-07,ButDepGor-PRA-12,MirSeiGor-JCTC-12}, yielding
\begin{equation}
	\label{eq:winftyspherical}
 w_\infty(r) =  \frac{1}{2(r+f(r))} - \frac{1}{2} v_H(r).
\end{equation}
In turn, $f(r)$ is a fully non-local functional of the density $\rho(r)$, given in terms of the cumulant $N_e(r)$ of Eq.~(\ref{eq:cumulant}) and its inverse $N_e^{-1}$,
\begin{equation}\label{eq:f}
f(r) = N_e^{-1}(2-N_e(r)).
\end{equation} 
In Fig.~\ref{fig:winfinity}, we report the energy densities $w_\infty(r)$ for the analytical solutions corresponding to the $\omega$ values of Table~\ref{tab:omega}.

\subsection{Energy densities at $\lambda=1$}
\begin{figure*}
  \includegraphics[width=0.5\textwidth]{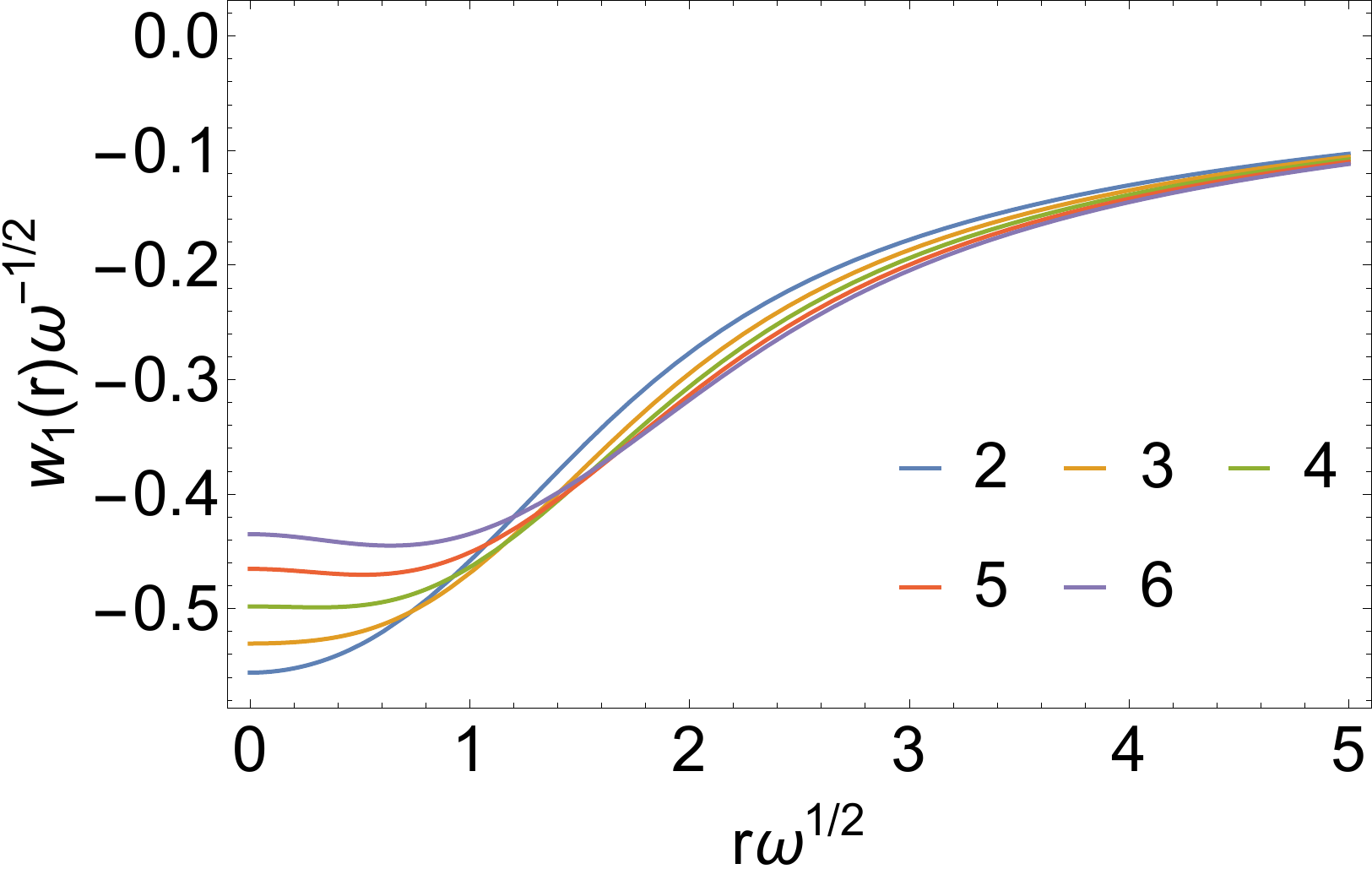}
\includegraphics[width=0.5\textwidth]{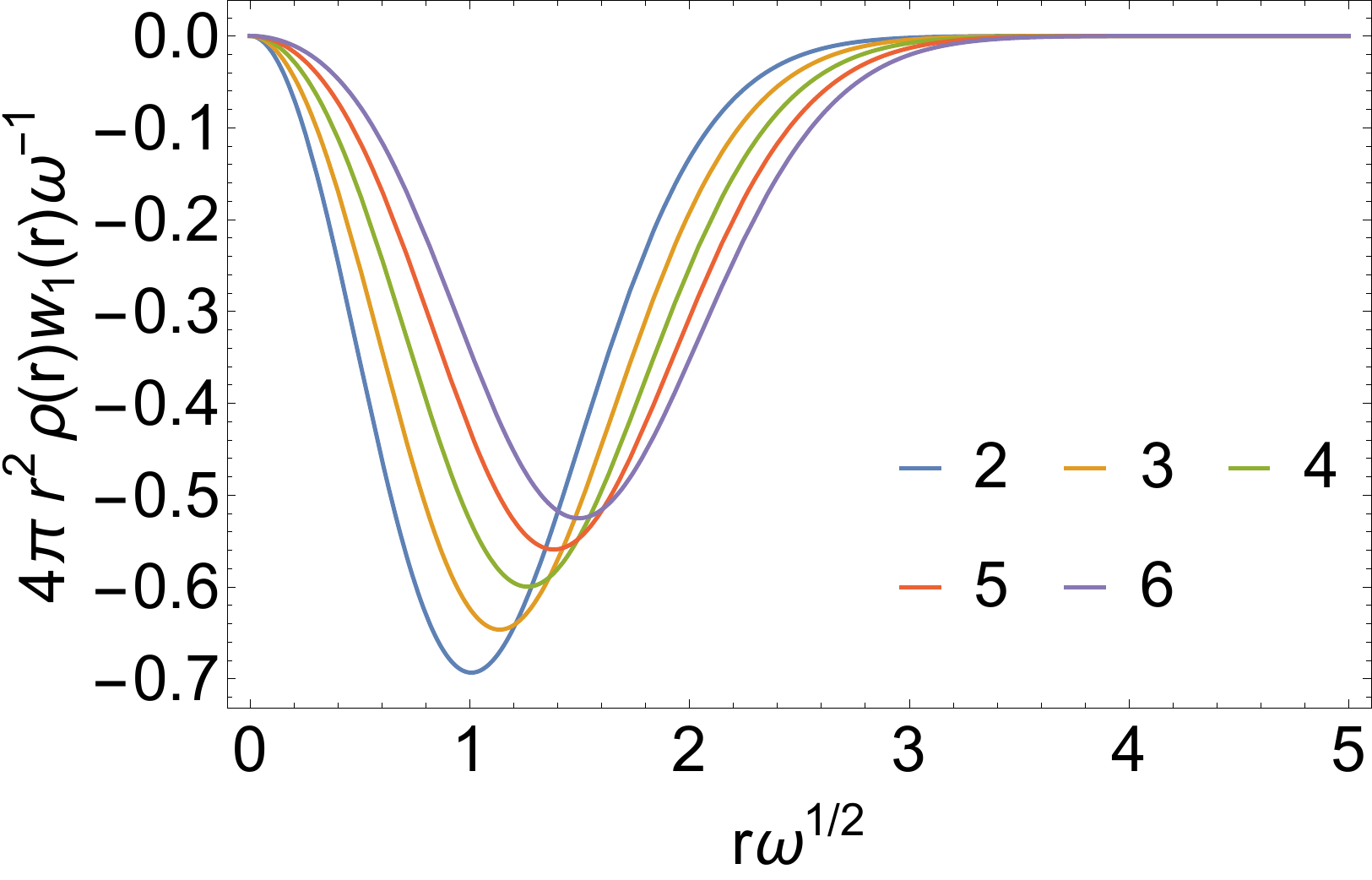}
\caption{Energy densities corresponding to $\lambda=1$ (first panel), and energy densities corresponding to $\lambda=1$ multiplied by the density and the volume element (second panel). The coordinates and energy densities are scaled according to the large $\omega$ limit.}
\label{fig:w1plot}       
\end{figure*}

\begin{figure*}
  \includegraphics[width=0.5\textwidth]{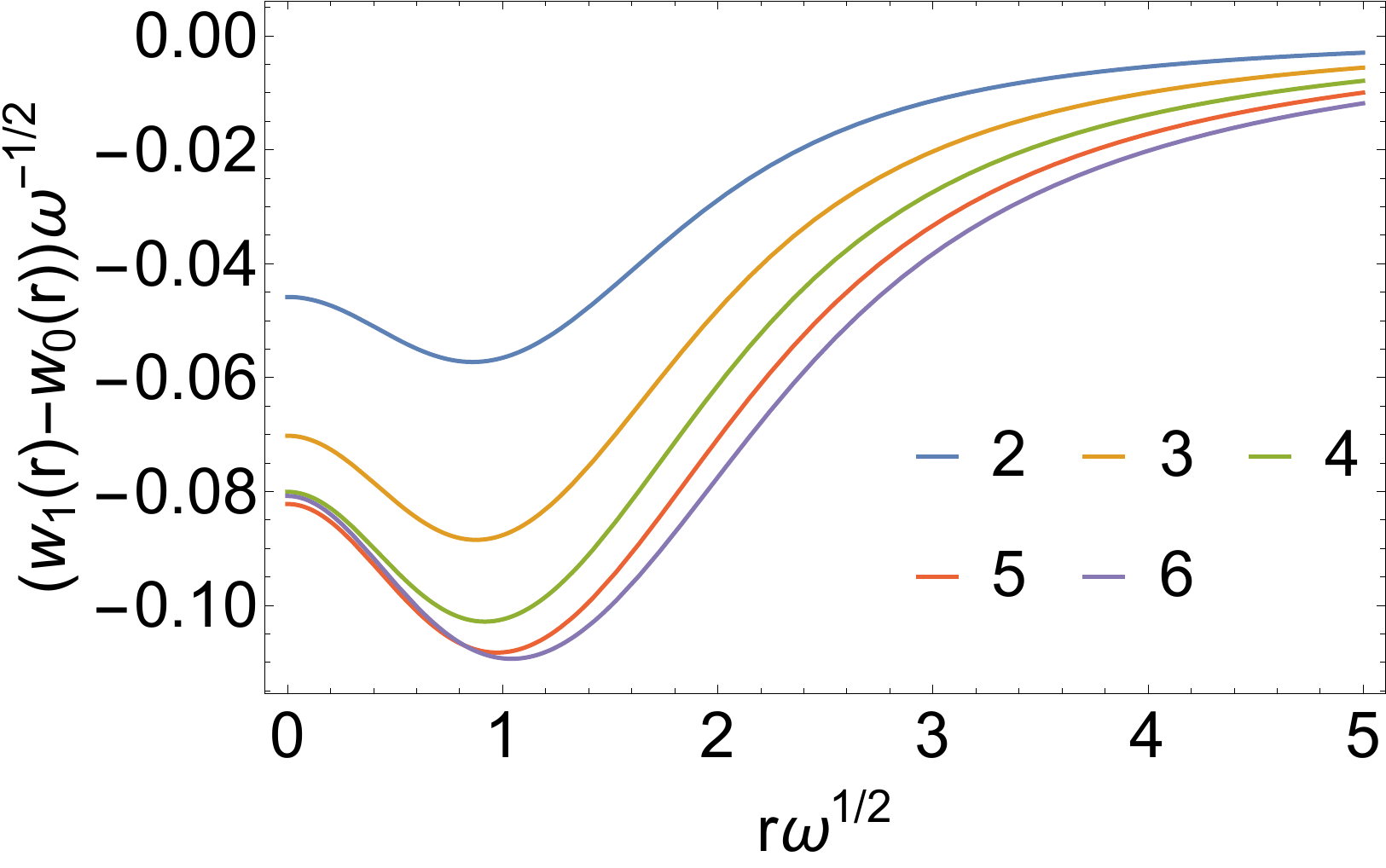}
\includegraphics[width=0.5\textwidth]{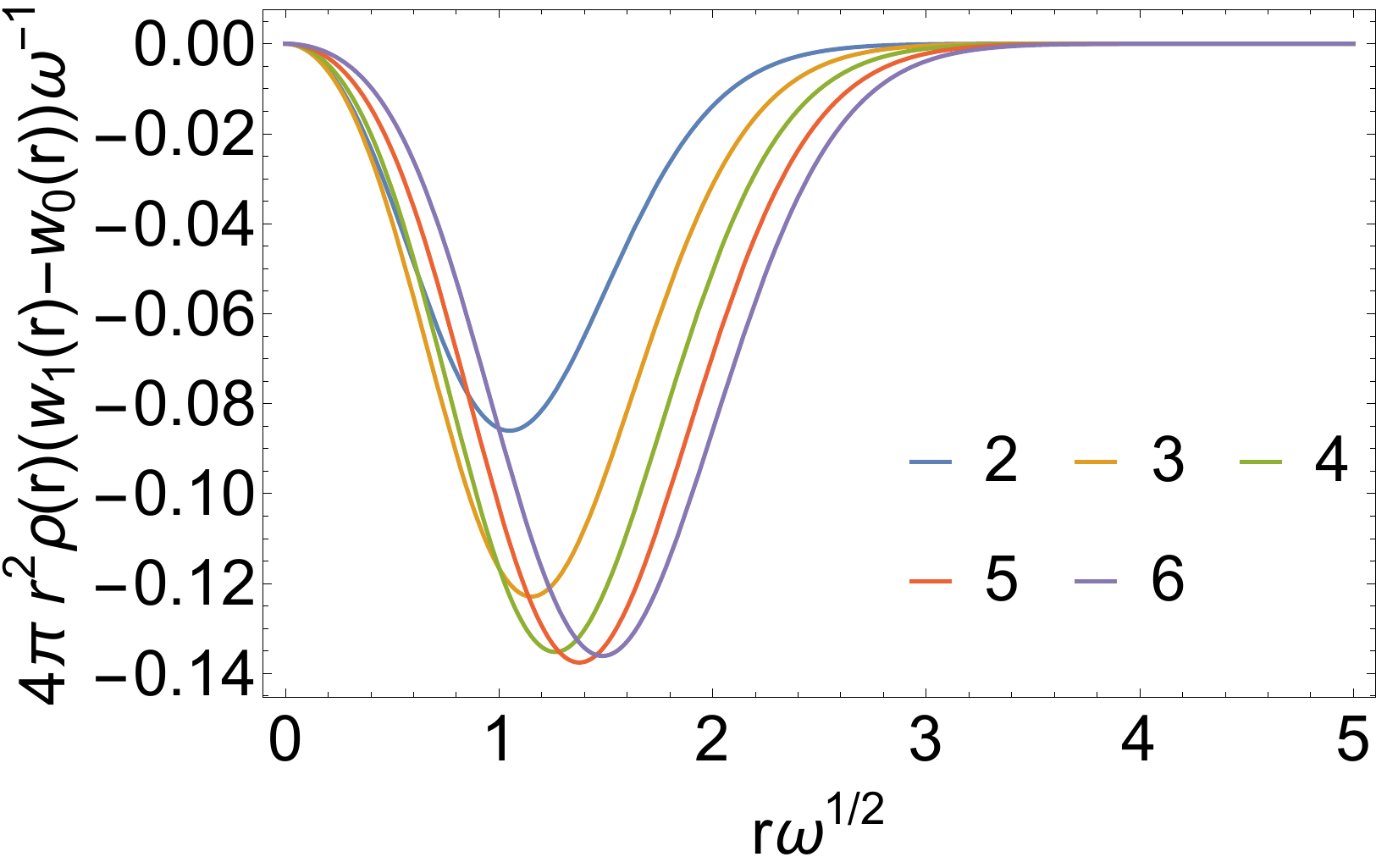}
\caption{Correlation energy densities (first panel) and correlation energy densities multiplied by the density and the volume element (second panel). The coordinate and energy density are scaled according to the large $\omega$ limit}
\label{fig:wcplot}       
\end{figure*}

Since we have exact analytic wavefunctions we can also compute the exact energy densities at physical coupling strength $\lambda=1$, which can be used to test the accuracy of local interpolations between $\lambda=0$ and $\lambda=\infty$, as well to study features of the energy densities as the interaction strength is changed. The exact $w_1(r)$ are reported in Fig.~\ref{fig:w1plot}. We see that the physical energy densities $w_1(r)$ for the Hooke's atom series differ more among each other at large $r$, unlike $w_0(r)$ and $w_\infty(r)$. This is clearer if we look at the correlation energy density $w_c(r) = w_1(r)-w_0(r)$, which is reported in Fig.~\ref{fig:wcplot}. The correlation energy density $w_c(r)$ decays $\propto -\frac{1}{r^3}$, but with different coefficients for different values of $\omega$.

\begin{figure*}
  \includegraphics[width=0.5\textwidth]{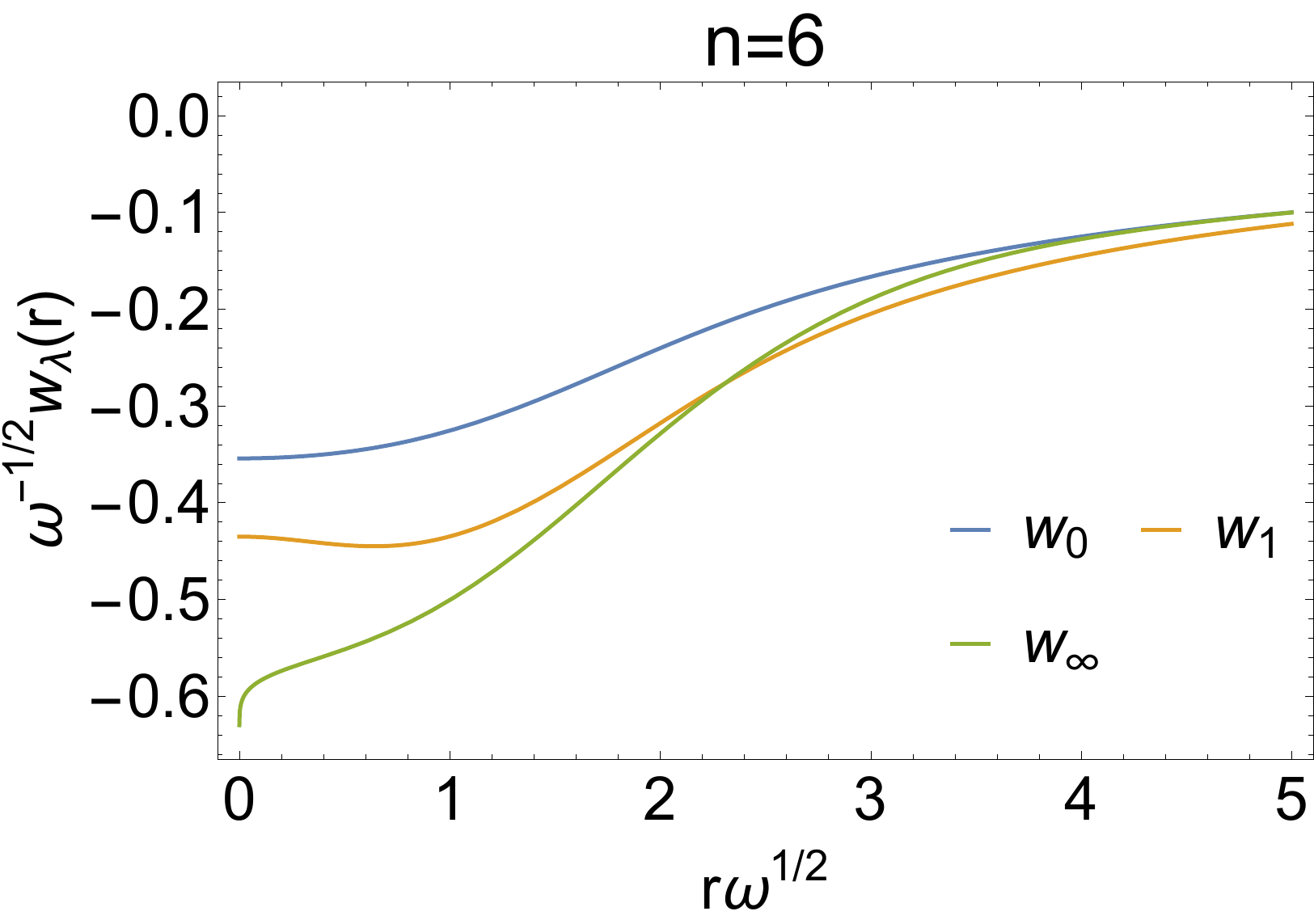}
\includegraphics[width=0.5\textwidth]{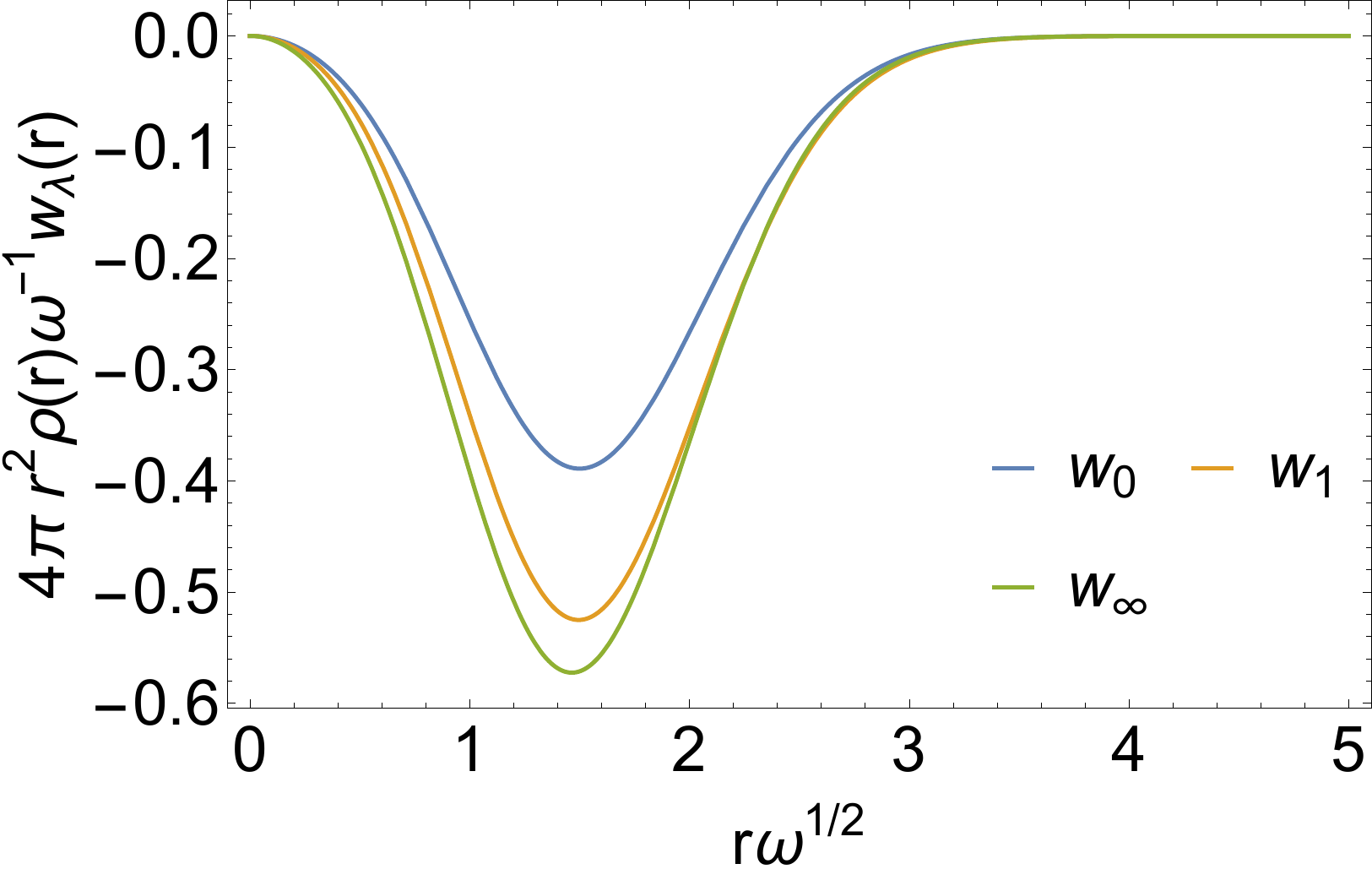}
\caption{Energy densities for the most strongly correlated Hooke's atom considered here ($n=6$), at different values of $\lambda$ (first panel). In the second panel the energy densities have been multiplied by the density and the volume element.}
\label{fig:n6plot}       
\end{figure*}

A comparison of the three energy densities $w_0$, $w_1$ and $w_\infty$ is given in Fig.~\ref{fig:n6plot} for the Hooke's atom with $n=6$. An interesting feature of these energy densities, already observed in Ref.~\cite{MirSeiGor-JCTC-12}, is that for large $r$ it can be seen that $w_1(r) < w_\infty(r)$, while for the corresponding global quantities we have the strict inequality $W_1[\rho] > W_\infty[\rho]$. However taking $w_1(r) \approx w_\infty(r)$ for large $r$ only has a small effect on the energy even for the most strongly correlated Hooke's atom considered here ($n=6$), as it becomes clear once the energy densities are multiplied by the density and the volume element (second panel of Fig~\ref{fig:n6plot}), which is what ultimately determines the correlation energy. This crossing of energy densities has never been observed, so far, in systems with the Coulomb external potential.

\section{Results from global and local interpolations}
\subsection{Interpolations using global ingredients}
The global ingredients $W_0[\rho]$, $W_0'[\rho]$  have been obtained as described in Sections~\ref{sec:w0Hooke} and \ref{sec:slopeHooke}, while $W_\infty[\rho]$ has been obtained by integrating the energy density of Eq.~(\ref{eq:winftyspherical}). Additionally, we have also obtained $W_\infty'[\rho]$ of Eq.~(\ref{eq_strong}), which in this case is given by \cite{GorVigSei-JCTC-09}
\begin{equation}
	W_\infty'[\rho]=\frac{1}{2}\int_0^\infty 4\pi\, r^2 \frac{\rho(r)}{2} \left(\omega_1(r)^2+\frac{\omega_2(r)^2}{2}\right)\,dr,
\end{equation}
with 
\begin{eqnarray}
	\omega_1(r)^2 & = & \frac{r^2+f(r)^2}{r f(r)(r+f(r))^3} \\
	\omega_2(r)^2 & = & -\frac{2(1+f'(r)^2)}{f'(r)(r+f(r))^3},
\end{eqnarray}
and with $f(r)$ given by Eq.~(\ref{eq:f}). Notice that $f'(r)<0$, so that $\omega_2(r)^2>0$. 

\begin{figure*}
  \includegraphics[width=0.5\textwidth]{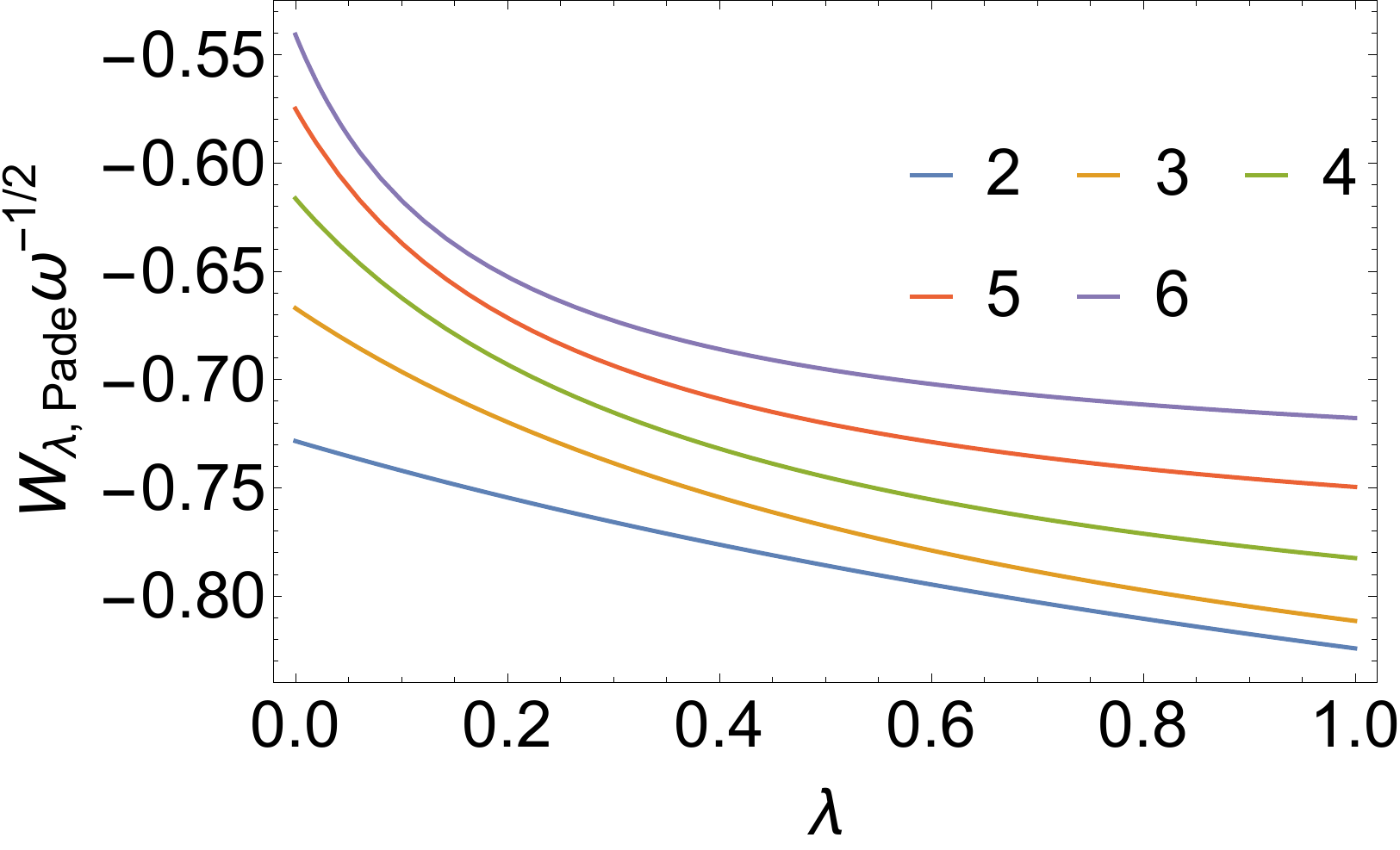}
\caption{The scaled adiabatic connection integrand as a function of $\lambda$ obtained from a Pad\'{e} interpolation that includes the exact $W_1[\rho]$ (see Appendix~\ref{app:formulas}).}
\label{fig:padeACplot}       
\end{figure*}

We have used the interpolation formulas reported in Appendix~\ref{app:formulas}, namely SPL \cite{SeiPerLev-PRA-99}, LB \cite{LiuBur-PRA-09}, ISI \cite{SeiPerKur-PRL-00} and revISI \cite{GorVigSei-JCTC-09}. The first two, SPL and LB, use only three ingredients (they do not include $W_\infty'[\rho]$), while ISI and revISI use all the four ingredients of Eqs.~(\ref{eq_weak})-(\ref{eq_strong}). Additionally, we have also used a Pad\'e approximant (see Appendix~\ref{app:formulas}) which uses $W_0[\rho], W_0'[\rho]$ and the exact $W_1[\rho]$, to generate plausible reference adiabatic connection curves, which are shown in Fig.~\ref{fig:padeACplot}.  As expected, as the Hooke's atoms get more correlated, the AC integrand displays a stronger curvature.

\begin{figure*}
  \includegraphics[width=0.6\textwidth]{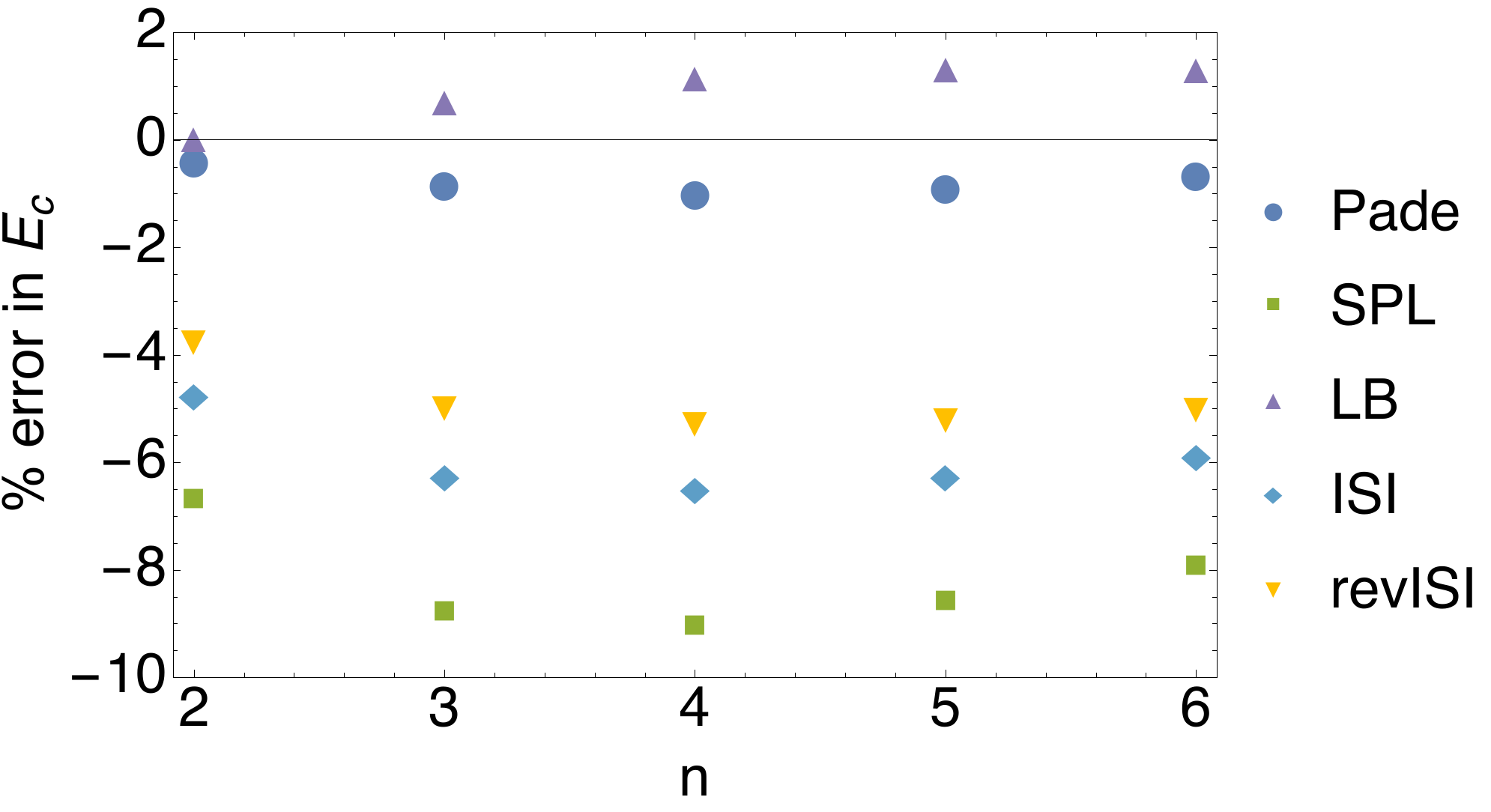}
\caption{Errors in the correlation energy resulting from the application of several global interpolations (see Appendix~\ref{app:formulas}).}
\label{fig:Ecglobalplot}       
\end{figure*}

The error resulting in the correlation energy $E_c[\rho]$ with the different global interpolations is shown in Fig.~\ref{fig:Ecglobalplot}. We consider only the correlation energy, since all the methods utilize 100\% exact exchange. The Pad\'{e} method performs best as expected, since it uses the exact $W_1$, which in practical situations is unavailable. The LB interpolation formula performs second best, while SPL, containing the same ingredients, performs much worse. The ISI and revISI methods improve slightly the SPL formula, but are still outperformed by LB, despite containing more exact information in the form of $W_\infty'[\rho]$.

\begin{figure*}
  \includegraphics[width=0.5\textwidth]{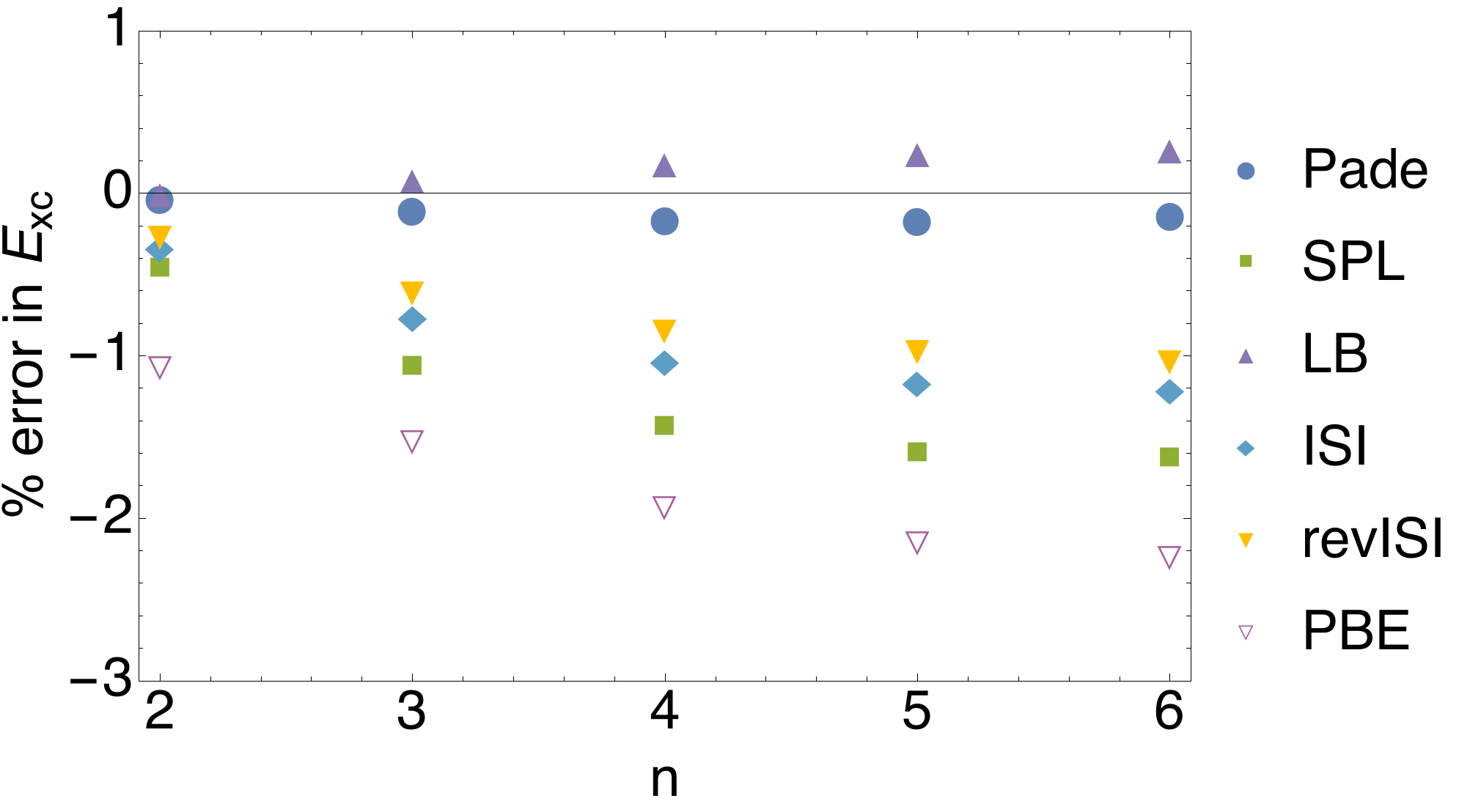}
\includegraphics[width=0.5\textwidth]{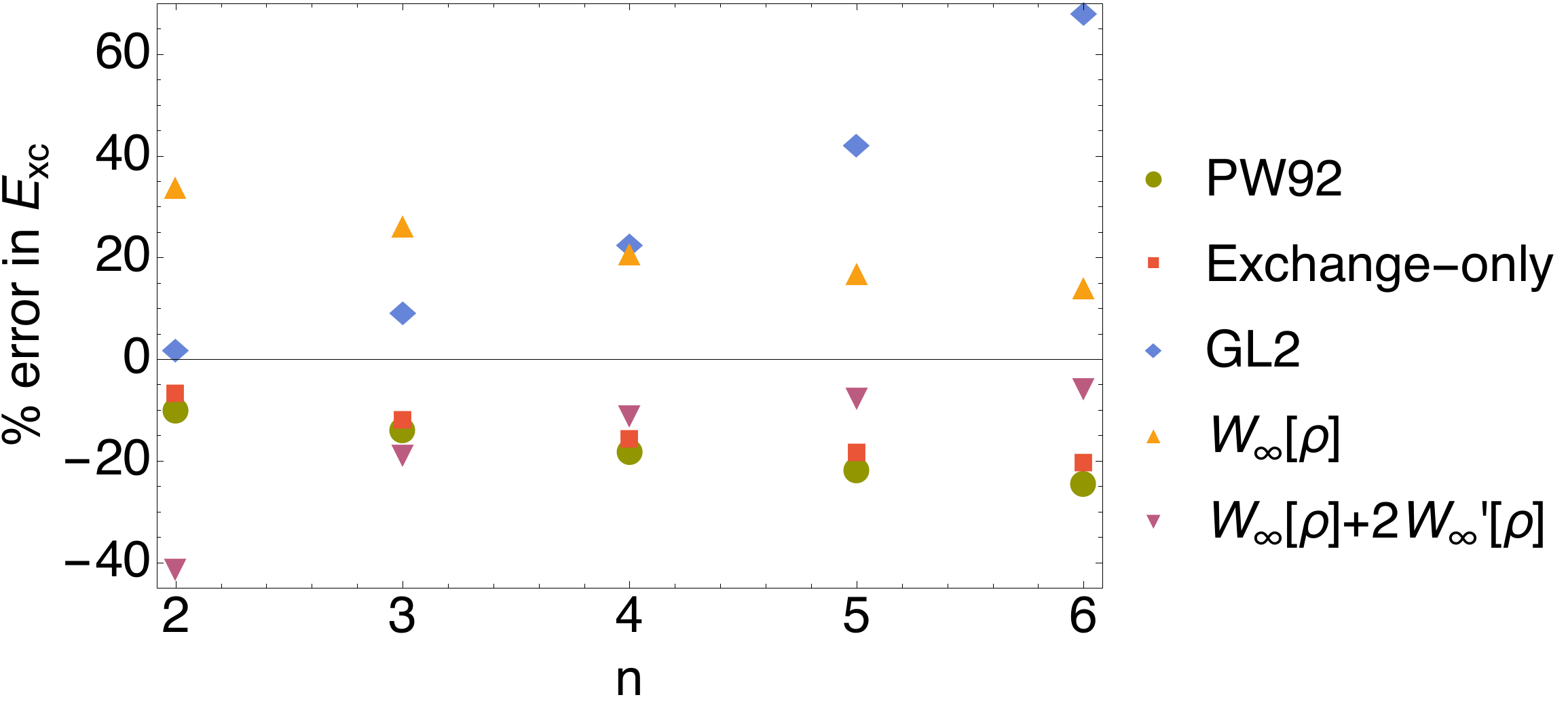}
\caption{Errors in the exchange-correlation energy resulting from the application of several global interpolations and approximations (see text).}
\label{fig:Excglobalplot}       
\end{figure*}

For comparison with traditional Density Functional Approximations (DFAs), such as the local density approximation (LDA) \cite{PerWan-PRB-92} and the PBE GGA \cite{PerBurErn-PRL-96}, we show the error in the exchange-correlation energy $E_{xc}[\rho]$ in the first panel of Fig.~\ref{fig:Excglobalplot}. It is clear that the adiabatic connection interpolation methods outperform the PBE method, however at the increased computational cost of a double hybrid. In the second panel of 
Fig.~\ref{fig:Excglobalplot} we compare the performance of LDA (PW92 \cite{PerWan-PRB-92}) with GL2 alone and with the $\lambda\to\infty$ expansion of Eq.~(\ref{eq_strong}) alone, which yields $E_{xc}[\rho]=W_\infty[\rho]$ if we retain only the first term, and $E_{xc}[\rho]=W_\infty[\rho]+2 W_\infty'[\rho]$, if we include also the second term. The LDA performs poorly already for the first Hooke's atom and its performance worsens as correlation increases. The GL2 method works well for the first Hooke's atom, which is expected since its adiabatic connection integrand resembles a straight line in  Fig.~\ref{fig:padeACplot}, but it is way too negative for the exchange-correlation energy in the more correlated Hooke's atoms. The $\lambda\to\infty$ expansion alone performs better as the Hooke's atoms become more correlated, but with the first term only is still too negative by about 15\% in the strongest correlated Hooke's atom. Adding the second term contribution reduces the error for $n>3$, and the resulting XC energy becomes now less negative than the exact one.

\subsection{Interpolations on energy densities}
\begin{figure*}
  \includegraphics[width=0.5\textwidth]{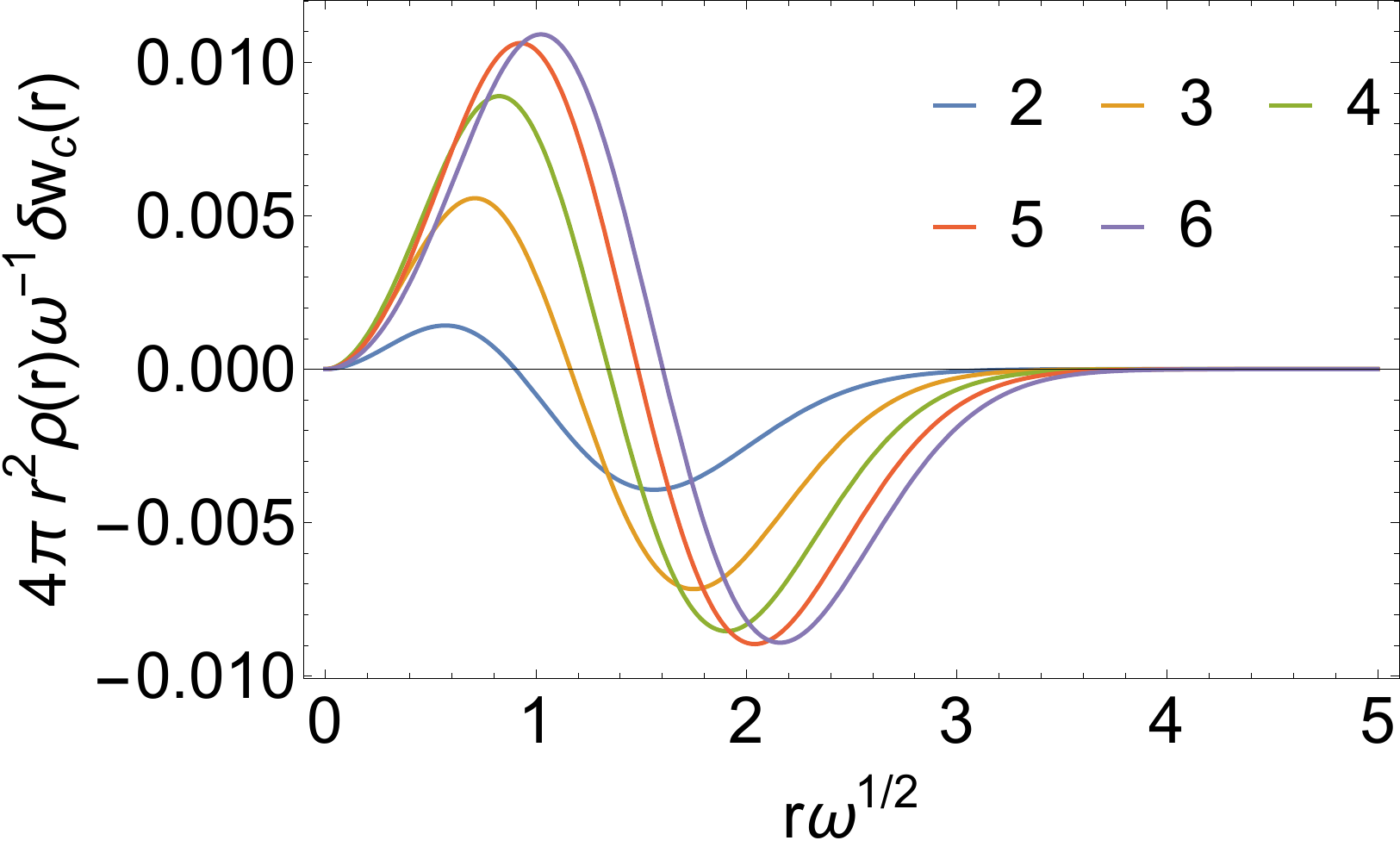}
\includegraphics[width=0.5\textwidth]{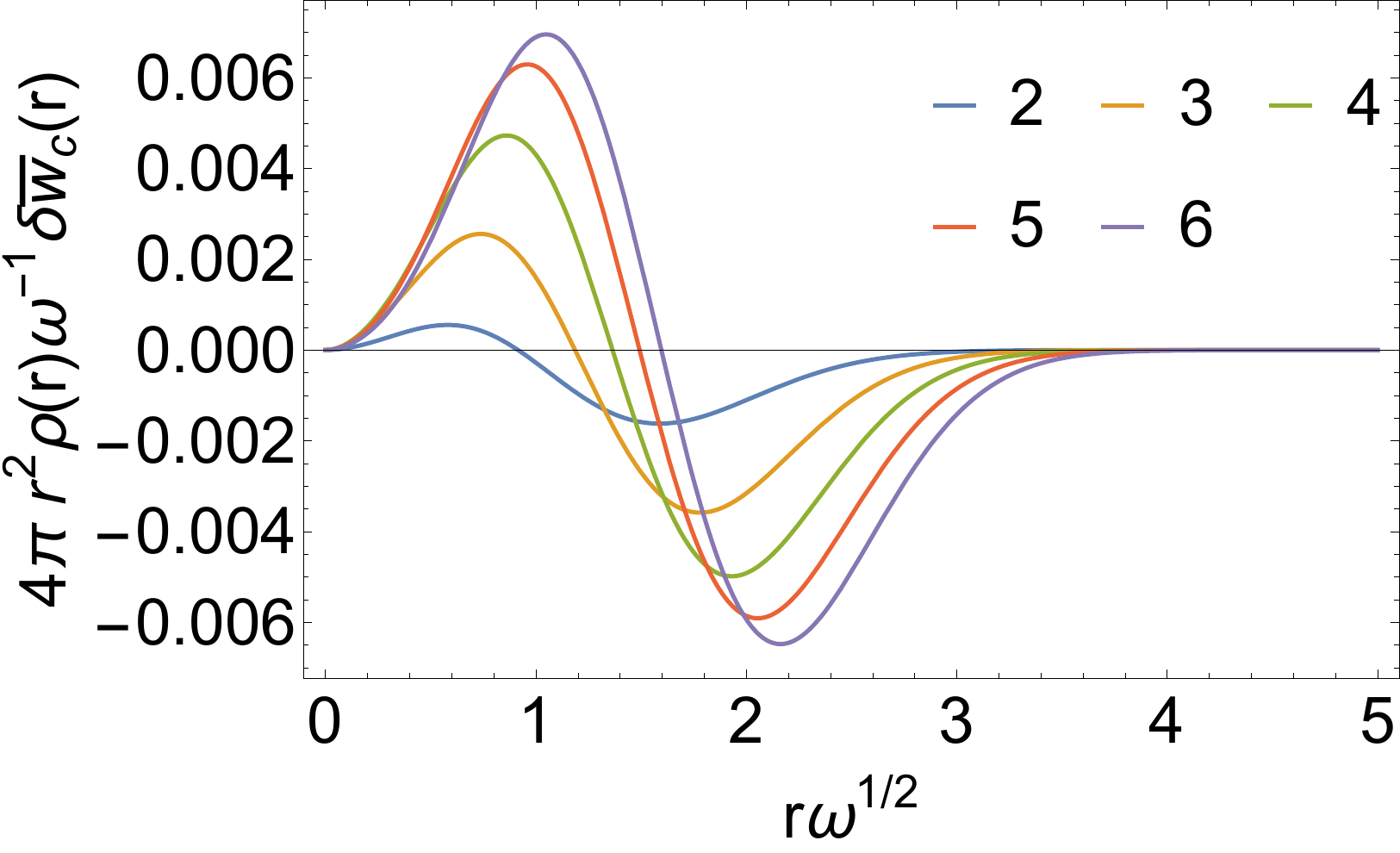}
\caption{Error $\delta w_c(\mathbf{r}) = w_{c, exact}(\mathbf{r})-w_{c, model}(\mathbf{r})$ multiplied by the volume element and density obtained with the LB approximation (first panel) and error in $\bar{w}_c(r)$ obtained with the same LB approximation (second panel). The high density scaling is applied. For the LB interpolation formula, see Appendix~\ref{app:formulas}}
\label{fig:burkeerrorplot}       
\end{figure*}

\begin{figure*}
  \includegraphics[width=0.5\textwidth]{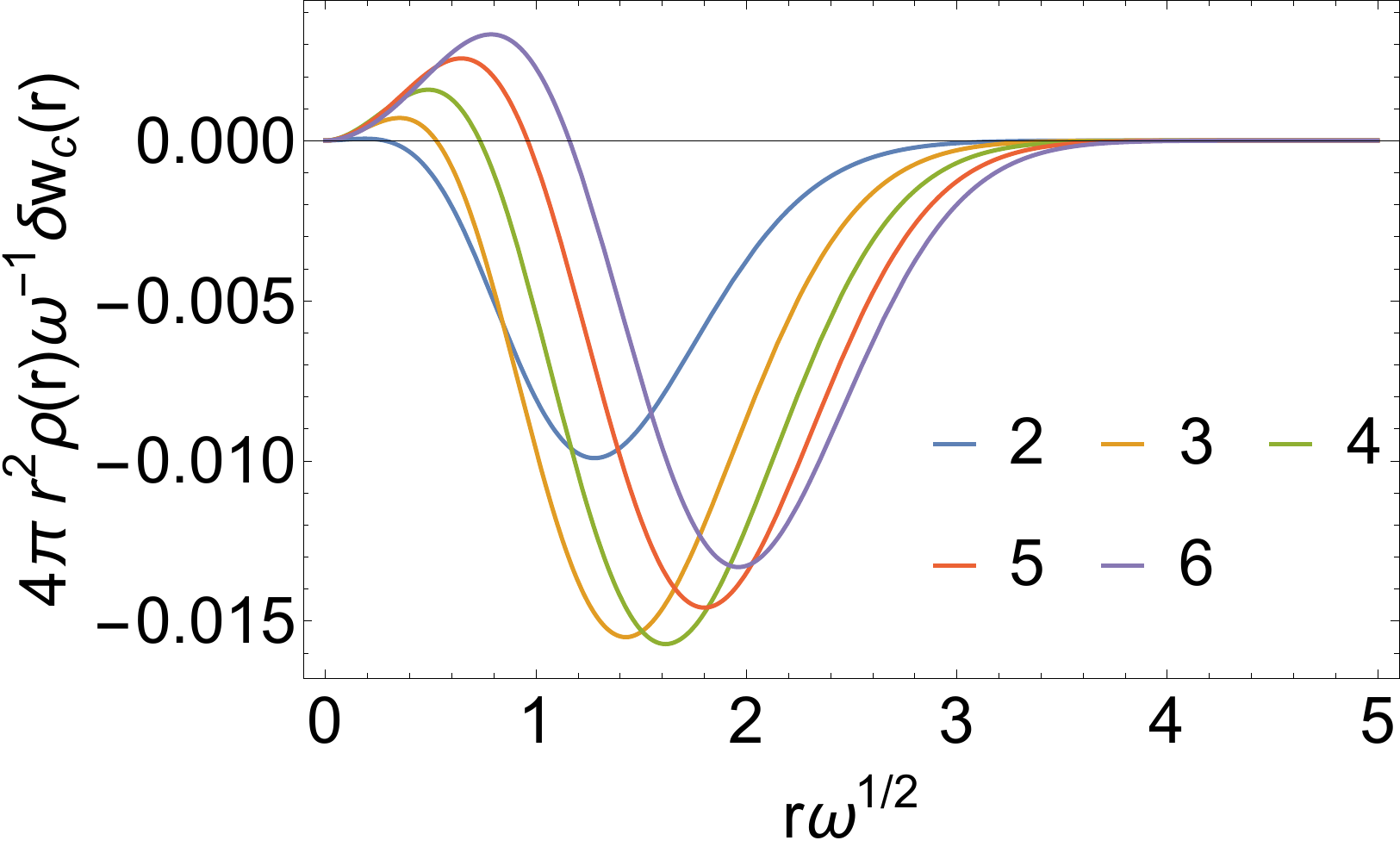}
\includegraphics[width=0.5\textwidth]{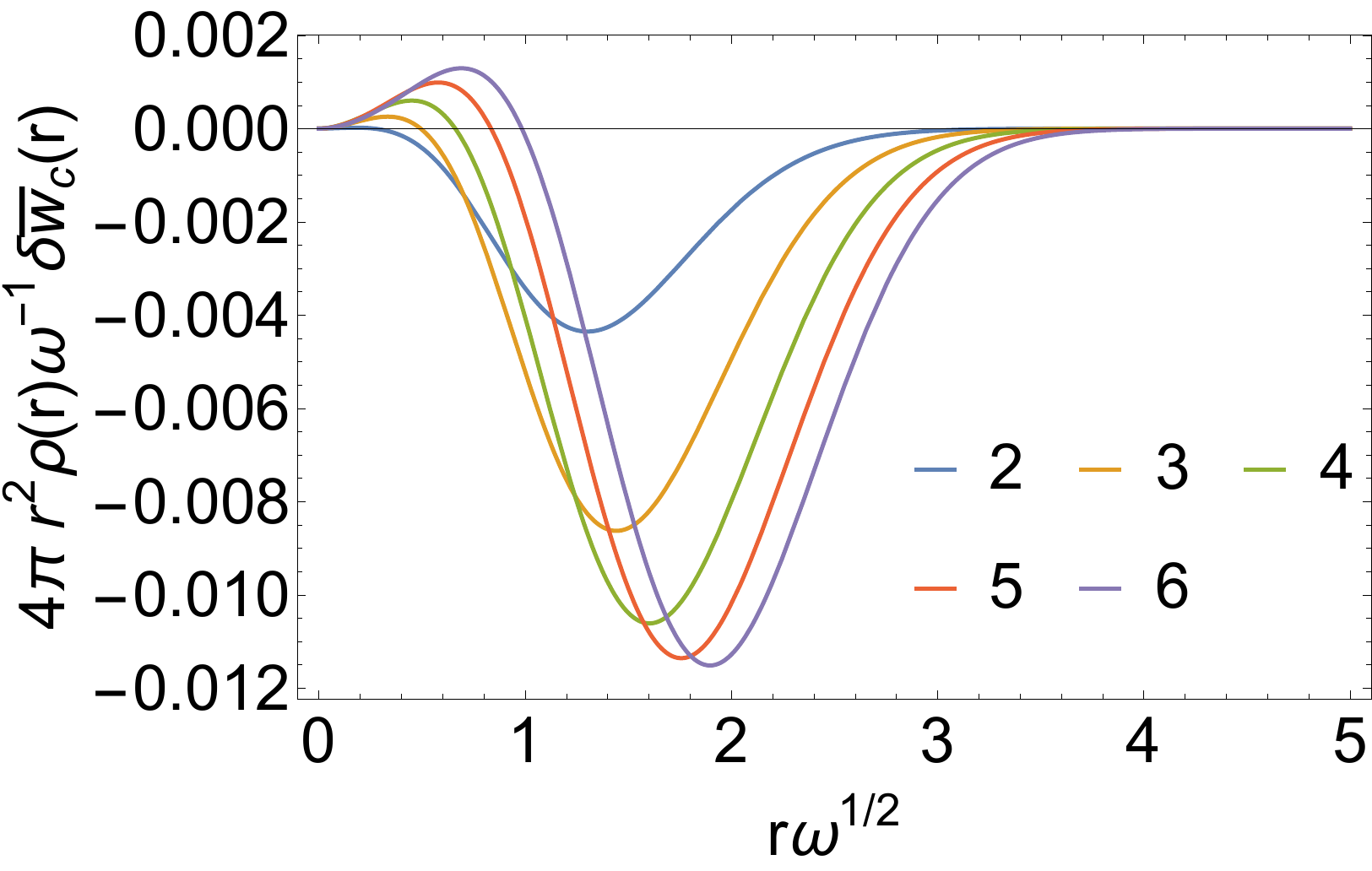}
\caption{Error $\delta w_c(\mathbf{r}) = w_{c, exact}(\mathbf{r})-w_{c, model}(\mathbf{r})$ multiplied by the volume element and density obtained with the SPL approximation (first panel) and error in $\bar{w}_c(r)$ obtained with the same SPL approximation (second panel). The high density scaling is applied. For the SPL interpolation formula, see Appendix~\ref{app:formulas}}
\label{fig:splerrorplot}       
\end{figure*}

As already mentioned at the end of Sec.~\ref{sec:theory}, an expression for the energy density corresponding to $W'_\infty[\rho]$ in the gauge of Eq.~(\ref{eq:energydef}) is not available. For this reason, we can only test local interpolations using the LB and SPL interpolation formulas, which do not use the information from $W'_\infty[\rho]$. We first compare the resulting $w_c(r)=w_1(r)-w_0(r)$ from the two interpolation formulas in the first panels of Figs.~\ref{fig:burkeerrorplot} (LB) and \ref{fig:splerrorplot} (SPL) with the exact result obtained from the analytic wavefunctions. The errors are small on an absolute scale, so we show in both figures $\delta w_c(\mathbf{r}) = w_{c, exact}(\mathbf{r})-w_{c, model}(\mathbf{r})$ and include the volume element and density. Notice that $\delta w_c(\mathbf{r}) = \delta w_1(\mathbf{r})$ since we use the exact $w_0(\mathbf{r})$ in the construction of both the LB and SPL approximations. In order to assess the coupling constant integrated energy density $\bar{w}_c$, which is not known exactly for any of the Hooke's atoms, we compare it with the one obtained from the Pad\'{e} interpolation, which includes the exact $w_0(r)$, $w_0'(r)$ and $w_1(r)$.

We see that in the case of LB there is an over-estimation of the coupling-constant averaged energy density at small $r$, which cancels quite well with an underestimation at large $r$, achieving almost perfect error cancellation. In the case of SPL, there is a smaller overestimation of the correlation at small $r$, coupled with a stronger underestimation of the correlation energy at large $r$, which worsens its performance.

\subsection{Comparison between global and local interpolations}
\begin{figure}
 \includegraphics[width=0.5\textwidth]{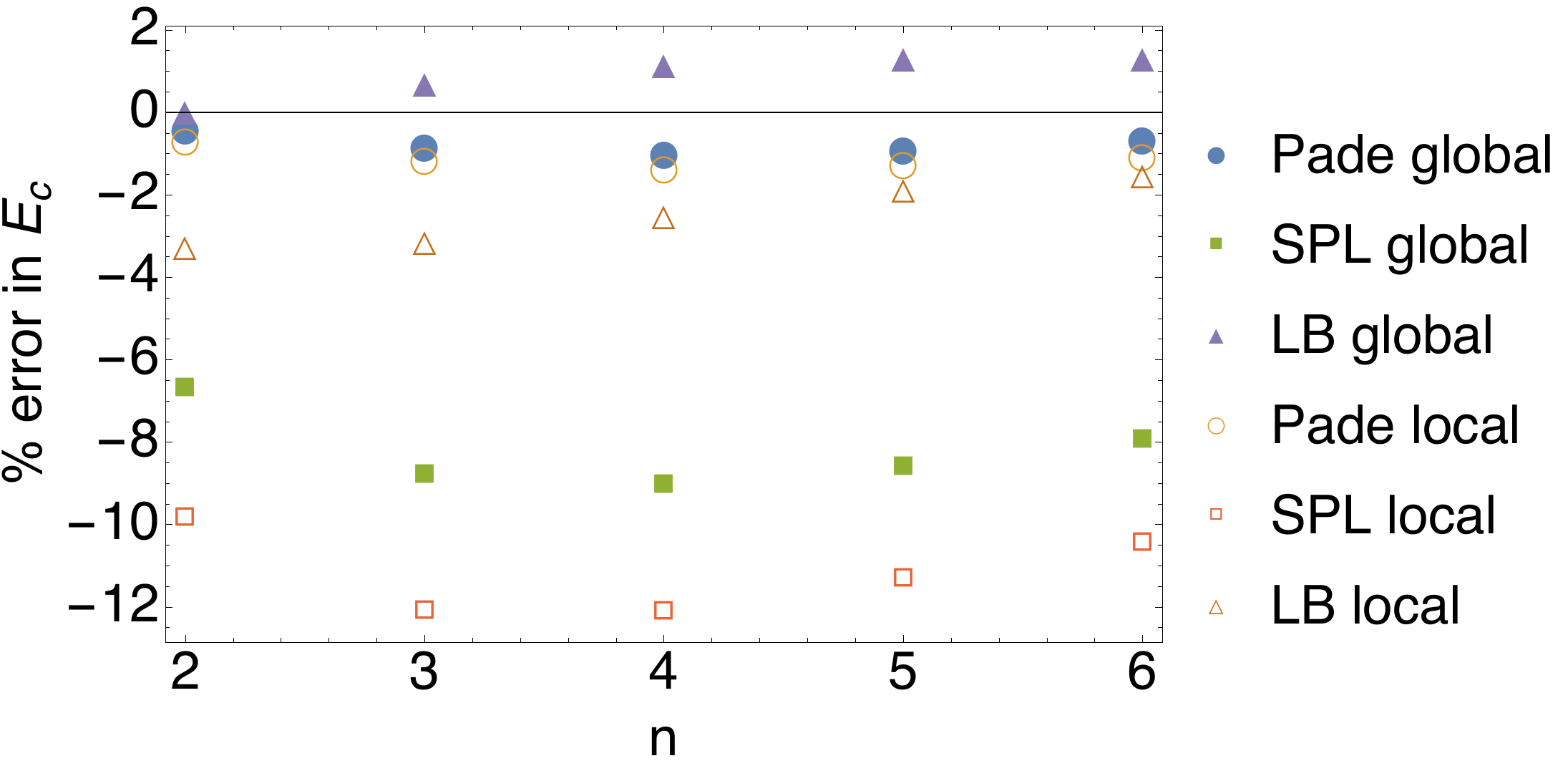}
\caption{Comparison of the local and global adiabatic interpolations in terms of the relative error in the correlation energy $E_c$.}
\label{fig:globlocplot}       
\end{figure}

Of interest is then comparing the performance of the global and local variants of the Pad\'{e}, LB and SPL interpolations. In Fig.~\ref{fig:globlocplot} the relative error on the correlation energy obtained from the local and global interpolation is shown, where in this case we use for both $10$ basis states per angular momentum quantum number for the slope. In the case of the Pad\'{e} interpolation the performance worsens only slightly going from the global to the local interpolation, while for the SPL interpolation there is a dramatic worsening. In the case of the LB interpolation the error switches sign for $n \geq 3$ and in general worsens. 

This is somehow surprising as, instead, for small chemical systems the local interpolations have been found to outperform their global counterparts \cite{VucIroWagTeaGor-PCCP-17,VucIroSavTeaGor-JCTC-16}.

\section{Kinetic correlation energy densities}
The coupling-constant integration is one possible way to recover the correlation part due to the difference between the true, interacting, kinetic energy $T[\rho]$ and the Kohn-Sham kinetic energy $T_s[\rho]$, $T_c[\rho]=T[\rho]-T_s[\rho]$. We have
\begin{equation}
	\label{eq:vkinbar}
	T_c[\rho]=\int \rho(\rv) (\overline{w}(\rv)-w_1(\rv))\,d\rv, 
\end{equation}
where $\overline{w}(\rv)$ is obtained by integrating $w_\lambda(\rv)$ over $\lambda$ between 0 and 1. Equation~(\ref{eq:vkinbar}) defines a possible kinetic correlation energy density equal to $\overline{w}(\rv)-w_1(\rv) $.

Another correlation kinetic energy density that has been defined \cite{BuiBaeSni-PRA-89} and studied \cite{BaeGri-PRA-96,GriLeeBae-JCP-96,BaeGri-JPCA-97} in the literature, and that has been found to display very interesting features for strongly correlated systems \cite{TemMarMai-JCTC-09,HelTokRub-JCP-09,YinBroLopVarGorLor-PRB-16,BenPro-PRA-16}, arises from the work of Baerends and coworkers \cite{BuiBaeSni-PRA-89,BaeGri-PRA-96,GriLeeBae-JCP-96,BaeGri-JPCA-97},
\begin{equation}
	\label{eq:vckin}
	v_{\rm c,kin}(\rv)=\frac{1}{2} \int \left( |\nabla_\rv\Phi(2,...,N|\rv)|^2  - |\nabla_\rv\Phi_s(2,...,N|\rv)|^2\right) \, d2..d N,
\end{equation}
where $\Phi(2,...,N|\rv)$ is a conditional amplitude defined in terms of a wavefunction $\Psi$ and its density $\rho$,
\begin{equation}
	\label{eq:condampl}
	\Phi(2,...,N|1)=\sqrt{\frac{N}{\rho(1)}}\Psi(1,...,N),
\end{equation}
$1,...N$ denote the spatial and spin coordinates of the $N$ electrons, and in Eq.~(\ref{eq:vckin}) we consider the conditional amplitude from the exact wavefunction (denoted with $\Phi$) and for the KS determinant (denoted with $\Phi_s$). Equation~(\ref{eq:vckin}) can also be rewritten in several different interesting and more practical forms, for example in terms of first order density matrices, or in terms of natural orbitals, or with Dyson orbitals (see, e.g., \cite{BaeGri-PRA-96,GriLeeBae-JCP-96,BaeGri-JPCA-97,RyaSta-JCP-14,CueAyeSta-JCP-15,CueSta-MP-16,KohPolSta-PCCP-16,GorGalBae-MP-16,RyaOspSta-JCP-17}). In the present case of $N=2$ electrons, Eq.~(\ref{eq:vckin}) takes the simple form
\begin{equation}
	\label{eq:vckinN2}
	v_{\rm c,kin}(r)=\frac{1}{2 \rho(r)} \int |\nabla_\rv\Psi(\rv,\rv')|^2 d\rv'  - \frac{|\nabla \rho(r)|^2}{8 \rho(r)^2},
\end{equation}
 where $\Psi(\rv_1,\rv_2)$ is the exact ground-state wavefunction of the interacting system.

Both $\overline{w}(\rv)-w_1(\rv)$ and $v_{\rm c,kin}(\rv)$ integrate to $T_c[\rho]$ when multiplied by the density $\rho(\rv)$, but they describe the kinetic correlation energy locally in a different way. Here we compare the features of these two definitions, as the correlation kinetic energy is important to capture strong correlation. Also, very recently, it has been proposed to use the correlated kinetic energy density as an additional variable in an extended KS DFT theory for lattice hamiltonians \cite{TheBucEicRugRub-arxiv-18}, and it is thus important to understand which definition is the most suitable to generalize this theory to the continuum. 

\begin{figure*}
  \includegraphics[width=0.5\textwidth]{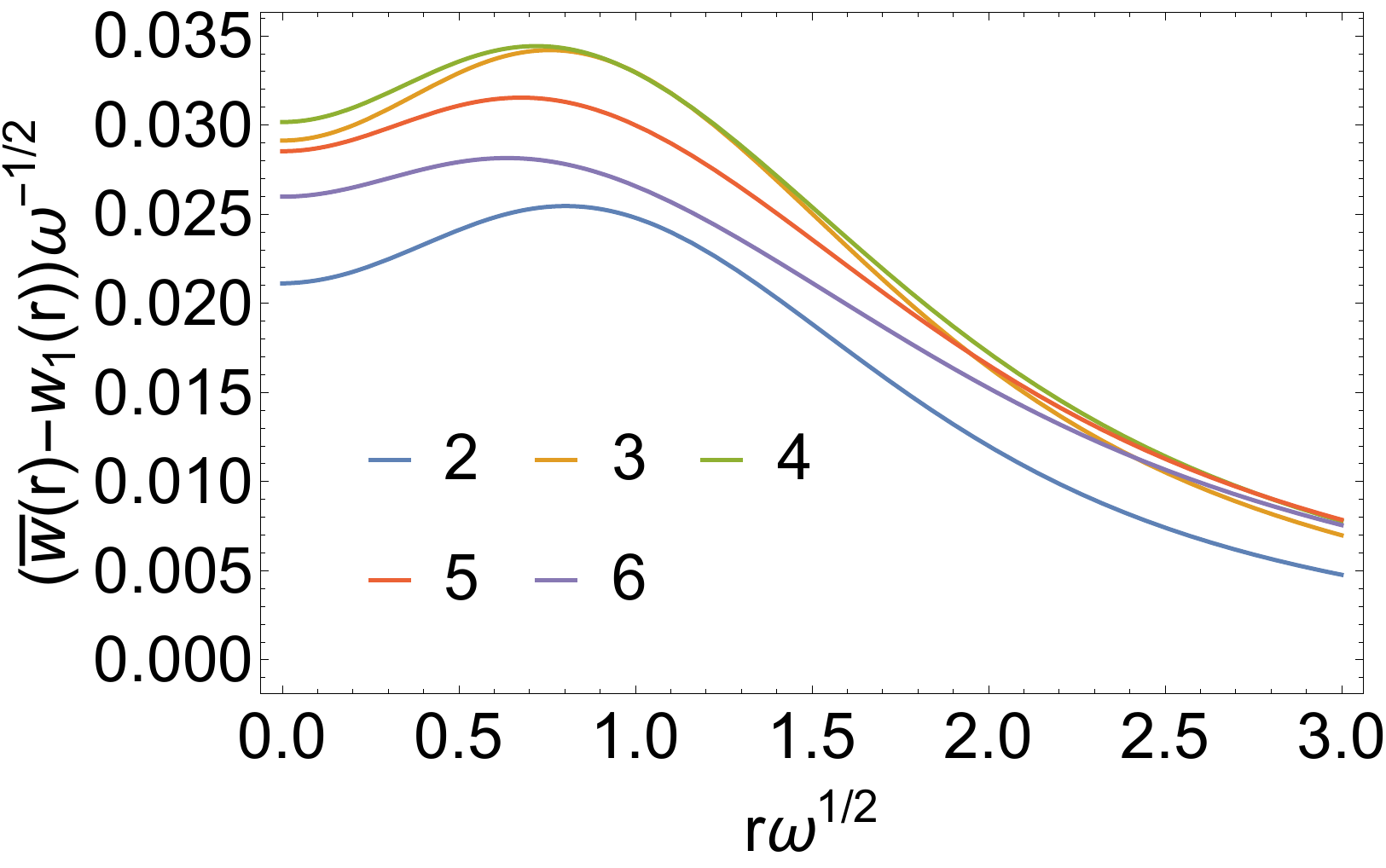}
\includegraphics[width=0.5\textwidth]{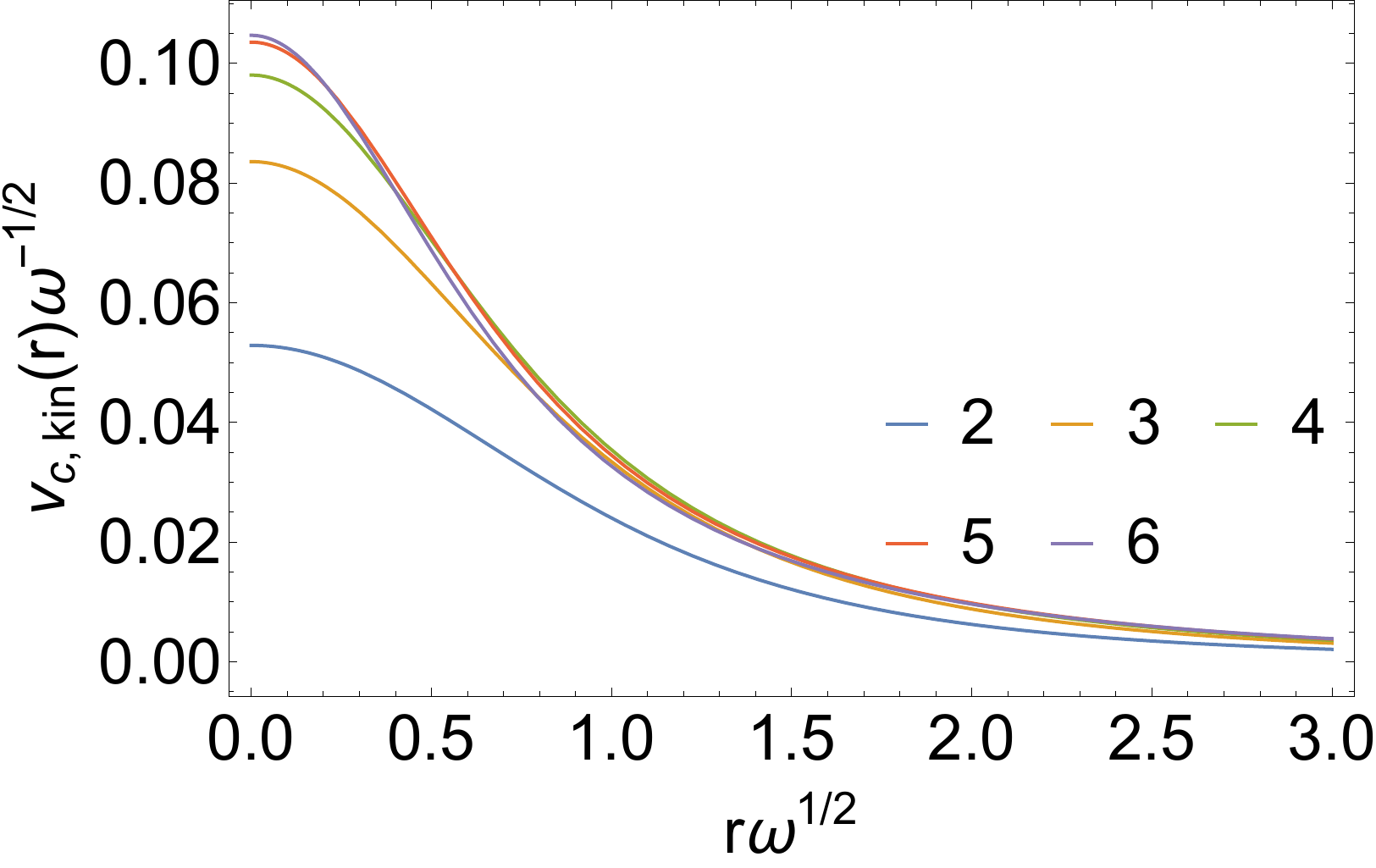}
\caption{ The two kinetic correlation energy densities of Eqs.~(\ref{eq:vkinbar}) and (\ref{eq:vckin}) for the different Hooke's atoms considered here. The high density scaling is applied.}
\label{fig:vckin}       
\end{figure*}

In Fig.~\ref{fig:vckin} we show the two different kinetic correlation energy densities, where for $\overline{w}(r)$ we have used the integration over $\lambda$ of the Pad\'e model, which uses the exact $w_0$, $w_0'$ and $w_1$ as input. We see that the two are rather different: $v_{\rm c,kin}(r)$ displays a peak in the center of the harmonic trap, reminiscent of the one appearing in a stretched bond \cite{BuiBaeSni-PRA-89,HelTokRub-JCP-09,YinBroLopVarGorLor-PRB-16,BenPro-PRA-16}, while $\overline{w}(\rv)-w_1(\rv)$ displays a weaker peak, which is not located at the center.

\subsection{Analysis of the peak of $v_{\rm c,kin}(\rv)$}
In the case of a stretched bond, it has been shown that the height of the peak of $v_{\rm c,kin}(\rv)$ at the midbond saturates as the bond is stretched \cite{HelTokRub-JCP-09}, displaying an anomalous scaling \cite{YinBroLopVarGorLor-PRB-16}, which is the way in which exact KS DFT can describe Mott-insulator physics \cite{YinBroLopVarGorLor-PRB-16}, and which is not captured by any approximate XC functional. In the low-density (small $\omega$ or large $n$) Hooke's atom, the system forms a ``Wigner molecule'', with the maximum of the density located away from the center of the harmonic trap. It is interesting to analyze how the height $v_{\rm c,kin}(0)$ of the peak scales when the system becomes very correlated ($\omega\to 0$), as in Fig.~\ref{fig:vckin} it seems to saturate when one uses the high-density scaling. 

\begin{figure}
  \includegraphics[width=0.5\textwidth]{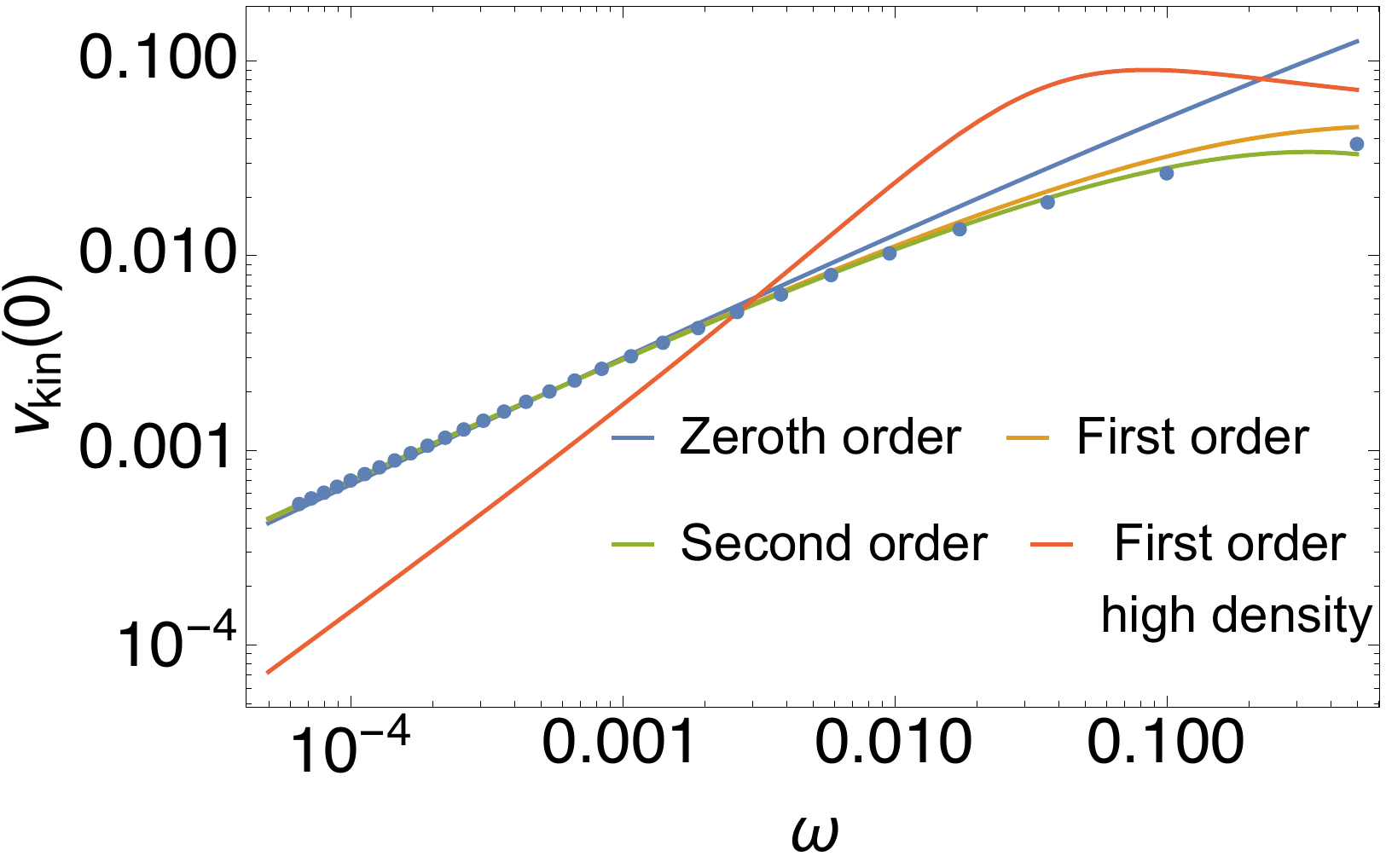}
\caption{The peak $v_{\rm c,kin}(0)$ as a function of $\omega$. The first three orders in the small-$\omega$ (strong correlation) expansion are compared with the values (dots) from the exact wavefunctions of Taut \cite{Tau-PRA-93}, and with the large-$\omega$ (weak correlation) expansion.}
\label{fig:peak}       
\end{figure}

For any 2-electron wavefunction of the form $\Psi(r_1,r_2,r_{12})=e^{-\frac{\omega}{2}(r_1^2+r_2^2)}p(r_{12})$, the peak's height is given by the simple expression
\begin{equation}
	\label{eq:peak}
		v_{\rm c,kin}(0)=\frac{\int_0^\infty  e^{-\omega x^2}\, x^2\, p'(x)^2 \,d x}{2 \int_0^\infty  e^{-\omega x^2}\, x^2\, p(x)^2 \,d x}.
\end{equation}
We have used up to the second-order of the small-$\omega$ (strong correlation) expansion of the exact wavefunction \cite{CioPer-JCP-00}, finding that in the scaling used in Fig.~\ref{fig:vckin} the peak does not saturate, but eventually will decrease and then go to zero very slowly, as $\omega^{1/6}$. In Fig.~\ref{fig:peak}, we show the peak's height as a function of $\omega$ for the analytic solutions, compared to the first three orders in the small-$\omega$ (strong correlation) expansion (Eq.~(32) of \cite{CioPer-JCP-00}), and with the large-$\omega$ (weak correlation) expansion (Eq.~(22) of \cite{CioPer-JCP-00}). We see that the strong-correlation expansion for the peak is much more accurate than ordinary perturbation theory from the weak correlation limit even for very moderate correlation (the Hooke's atom with $\omega=1/2$ resembles the He atom as far as the degree of correlation is concerned).

\section{Conclusions}
We have analyzed the performances of exchange-correlation functionals built from global and local interpolations between the weak- and the strong-interaction limits of DFT for the Hooke's atom series. This case study allows for the use of exact analytical input ingredients, thus disentangling the errors coming from the interpolation itself from those on the input quantities. Surprisingly, we have found that for these systems the global interpolations always outperform their local counterparts, in striking contrast with what had been observed so far for small chemical systems \cite{VucIroSavTeaGor-JCTC-16,VucIroWagTeaGor-PCCP-17}.

We have also compared two different definitions of the kinetic correlation energy density, which plays a crucial role for strongly correlated systems \cite{HelTokRub-JCP-09,YinBroLopVarGorLor-PRB-16}, and that can help in understanding how to extend to the continuum a KS theory that recovers the exact kinetic energy density recently proposed for lattice models \cite{TheBucEicRugRub-arxiv-18}.

\begin{acknowledgements}
Financial support from European Research Council under H2020/ERC Consolidator Grant corr-DFT (Grant Number 648932) is acknowledged. We thank S. Giarusso and S. Vuckovic for insightful discussions.
\end{acknowledgements}

\appendix
\section{Interpolation Formulas} \label{app:formulas}
In the following we report the interpolation formulas in terms of the global ingredients $W_0$, $W_0'$, $W_\infty$ and $W_\infty'$. For the interpolation on energy densities, we have used the same SPL, LB and Pad\'e[1/1] formulas below in each point of space, replacing the global quantities $W_i$ with their local counterparts $w_i(\rv)$. \\

\noindent{\bf  Interaction Strength Interpolation (ISI) formula} \cite{SeiPerKur-PRL-00,SeiPerKur-PRA-00}\\
\begin{equation}
W_\lambda^\ISI  = W_\infty + \frac{X }{\sqrt{1+\lambda Y }+Z }\ ,
\end{equation}
with
\begin{eqnarray}\label{Y}
&&X=\frac{xy^2}{z^2}\; ,\; Y=\frac{x^2y^2}{z^4}\; , \; Z=\frac{xy^2}{z^3}-1\ ;\\
&& x=-2 W_0' ,\; y=W_\infty'\; , \; z=W_0-W_\infty\ .
\end{eqnarray}
After integration in Eq.~(\ref{eq:Excadiab}), we have
\begin{equation}
E_{xc}^\ISI = W_\infty + \frac{2X}{Y}\left[\sqrt{1+Y}-1-Z\ln\left(\frac{\sqrt{1+Y}+Z}{1+Z}\right)\right]\ .
\end{equation}

\noindent{\bf Revised ISI (revISI) formula} \cite{GorVigSei-JCTC-09}\\ 
\begin{equation}
W_\lambda^\revISI  = W_\infty  + \frac{b  \left( 2 + c  \lambda + 2 d  \sqrt{1 + c \lambda}\right)}{2 \sqrt{1 + c \lambda} \left( d  + \sqrt{1 + c \lambda}\right) ^2} ,
\end{equation}
where
\begin{eqnarray}\label{c}
\nonumber
b & = &-\frac{4 W_0' (W_\infty')  ^{2}}{\left(W_0-W_\infty\right)^2}\; ,\; c=\frac{4 (W_0' W_\infty ')^2}{\left(W_0-W_\infty\right)^4}\; ,\; \\
d & = & -1-\frac{4 W_0' (W_\infty ')  ^{2}}{\left(W_0-W_\infty\right)^3} \ .
\end{eqnarray}
The corresponding XC functional is
\begin{equation}
E_{xc}^\revISI = W_\infty + \frac{b}{\sqrt{1+c}+d}\ .
\end{equation}

\noindent{\bf Seidl-Perdew-Levy (SPL) formula} \cite{SeiPerLev-PRA-99}\\
\begin{equation}\label{spl_eq}
W_\lambda^\SPL  = W_\infty  +\frac{W_0 -W_\infty }{\sqrt{1+2\lambda \chi }}\ ,
\end{equation}
with
\begin{equation}
\chi = \frac{W_0'}{W_\infty-W_0}\ .
\end{equation}
The SPL XC functional reads
\begin{equation}
E_{xc}^\SPL = \left(W_0-W_\infty\right)\left[\frac{\sqrt{1+2\chi}-1-\chi}{\chi}\right] + W_0\ .
\end{equation}
Notice that this functional does not make use of the information from $W_\infty'$.\\

\noindent{\bf Liu-Burke (LB) formula} \cite{LiuBur-PRA-09}\\
\begin{equation}
W_\lambda^\LB  = W_\infty  + \beta (y  + y^4 )\ , 
\end{equation}
where 
\begin{equation}
y = \frac{1}{\sqrt{1+\gamma\lambda}}\; , \; \beta=\frac{W_0-W_\infty}{2}\; , \; \gamma=\frac{4 W_0'}{5(W_\infty-W_0)}\ .
\end{equation}
Using Eq.~(\ref{eq:Excadiab}), the LB XC functional is found to be
\begin{equation}
E_{xc}^\LB = W_0 +2\beta\left[\frac{1}{\gamma}\left(\sqrt{1+\gamma}-\frac{1+\gamma/2}{1+\gamma}\right)-1\right]\ .
\end{equation}
Also the LB functional does not use the information from $W_\infty'$. \\

\noindent{\bf Pad\'e[1/1] formula with the exact $W_1$} \cite{Ern-CPL-96}\\
\begin{equation}
	W_\lambda^{\rm Pade} = a + \frac{b\, \lambda}{1+ c\, \lambda},
\end{equation}
with
\begin{eqnarray}
	a & = & W_0 \\
	b & = &  W_0' \\
	c & = &  \frac{W_1-W_0 - W_0'}{W_0-W_1},
\end{eqnarray}
yielding
\begin{equation}
E_{xc}^{\rm Pade}=a+b\,\left(\frac{c-\log(1+c)}{c^2}\right)
\end{equation}

\bibliographystyle{spphys}       

\end{document}